\RequirePackage{amsmath}
\documentclass[12pt]{iopart}
\usepackage{iopams} 
\usepackage{color}
\usepackage{graphicx}

\newtheorem{theorem}{Theorem}[section]
\newtheorem{definition}{Definition}[section]

\begin{document}

\title{Sachs equations for light bundles in a cold plasma}

\author{Karen Schulze-Koops$^{1,2}$, Volker Perlick$^2$ and Dominik J. Schwarz$^1$}

\address{${}^1$ Fakult{\"a}t f{\"u}r Physik, Universit{\"a}t Bielefeld, Postfach 100131, 33501 Bielefeld, Germany 
\\
${}^2$ ZARM, University of Bremen, 28359 Bremen, Germany. 
\\
Email:  volker.perlick@zarm.uni-bremen.de, dschwarz@physik.uni-bielefeld.de}
\vspace{10pt}

\begin{abstract}
We study the propagation of light bundles in non-empty spacetime, as most of the Universe is filled by baryonic matter 
in the form of a (dilute) plasma. Here we restrict to the case of a cold (i.e., pressureless) and non-magnetised plasma. 
Then the influence of the medium on the light rays is encoded in the spacetime dependent plasma frequency. 
Our result for a general spacetime generalises the Sachs equations to the case of a cold plasma Universe. 
We find that the reciprocity law (Etherington theorem), the relation that connects area 
distance with luminosity distance, is modified. Einstein's field equation is not used, i.e., our 
results apply independently of whether or not the plasma is self-gravitating. As an example, our findings 
are applied to a homogeneous plasma in a Robertson-Walker spacetime. We find small modifications of the 
cosmological redshift of frequencies and of the Hubble law.  
\end{abstract}

%
%
%
%
%

\section{Introduction}

For most applications of general relativity to astrophysics light propagation 
may be modelled in terms of rays and the influence of a medium may be neglected.
Then light rays are lightlike geodesics of the spacetime metric. 
However, the Universe is filled by a dilute medium which affects the propagation of light 
in various ways. Examples are dust grains, gas and molecular clouds, and most commonly plasma.
Here we focus on the effects of a plasma on the propagation of bundles of light rays and in the 
context of general relativity. While plasma effects on the propagation\footnote{We do not consider effects 
like Thomson scattering of visible light in a dilute cold plasma, by which only a small number of individual 
photons are randomly scattered out of or into a light bundle.} of visible light are negligible,
plasmas do effect radio waves. Here are two examples. Firstly, 
the travel time of a radio signal from a pulsar to the observer on the Earth is influenced 
by the interstellar medium. This is usually expressed in terms of the 
so-called \emph{dispersion measure} which is one of the most important observables 
in pulsar astronomy, see e.g.~\cite{LorimerKramer,LyneGraham2005}. Secondly,
the Solar corona has an effect on the travel time and on the bending angle of radio
signals; the relevant formulas, using the linearised version of gravity, have been
derived by Muhleman et al. \cite{MuhlemanJohnston1966,MuhlemanEkersFomalont1970}. 

In these two examples, and in most other applications to astrophysics, the medium may 
be modelled as a non-magnetised pressureless (``cold'') plasma. Then the influence 
of the medium onto the light rays is determined by the \emph{plasma frequency} 
$\omega _p$ which is related to the electron number density $n_e$ by
\begin{eqnarray} \label{eq:fp}
\omega_p (x) = \sqrt{\frac{e^2 n_e(x)}{m_e}}, \quad 
f_p (x) = \frac{\omega_p (x)}{2\pi} = 8980 \, \sqrt{n_e (x)}\,  {\rm Hz\, cm}^{3/2} \, .
\end{eqnarray}
Here $e$ is the electric charge of the electron, and $m_e$ is the mass of the electron.
Plasma effects modify the vacuum results by terms of order $(\omega_p (x)/\omega (x))^2$ 
and derivatives of that ratio, where $\omega (x)$ is the observed frequency with respect
to a chosen observer. If $\omega$ is much bigger than $\omega _p$, the 
influence of the medium is negligible, i.e., light rays are lightlike geodesics 
of the spacetime metric. If $\omega$ is bigger than $\omega _p$ but not much bigger, 
the influence of the plasma is palpable. If $\omega$ is lower than $\omega _p$, light cannot 
travel through the plasma at all. In all applications to astrophysics and cosmology,
the plasma frequency is in the radio regime. E.g., in the Solar corona the electron 
density drops from $\sim 10^8$ cm$^{-3}$ at the solar limb to $\sim 10^3$ cm$^{-3}$ at $10 \, R_\odot$, 
which corresponds to $f_p$ varying between $\sim 100$ MHz and $\sim 0.1$ MHz 
\cite{SolarCorona1,SolarCorona2,SolarCorona3}. In the interstellar space the plasma 
frequency is typically a few kHz, while in the intergalactic space it can be as low 
as a few Hz. This is the reason why effects from plasmas may be safely neglected 
in the optical frequency range. 

Without a low frequency radio telescope in a space orbit or on the Moon all radio signals 
from the Universe must pass Earth's ionosphere in which $n_e \sim 10^6$ cm$^{-3}$. Thus 
Earth's ionosphere is not transparent for frequencies below 10 MHz. Low frequency 
radio waves ($10$ MHz to $300$ MHz) are currently observed with modern radio interferometers, such as 
the Giant Metrewave Radio Telescope (GMRT) \cite{GMRT} and the Low Frequency 
Array (LOFAR) \cite{vanHaarlem}. Both are multi-purpose observatories. There are also several instruments 
dedicated to the detection of the 21 cm line of hydrogen and its intensity fluctuation at high redshifts, i.e. from the 
epochs of reionisation and cosmic dawn, such as the Murchison Widefield Array (MWA) \cite{MWA}. 
The best imaging resolution at the lowest frequencies, the aspect most relevant to this work, is currently obtained by 
means of LOFAR. The upgrade LOFAR2.0 will lead to further increase of sensitivity.
The next generation of instruments will include a low frequency instrument  of the Square Kilometre Array 
(SKA-LOW), which will improve todays sensitivities and survey speeds  
at frequencies between $50$ and $300$ MHz.  The space-based array 
Orbiting Low Frequency Antenna for Radio astronomy (OLFAR), using a swarm of nano-satellites in 
Moon orbits, has been suggested \cite{BudianuMeijerinkBentum2015,BentumBonettiSpallicci2017}. 
This is supposed to operate in the frequency range of 0.1 to 15 MHz, so 
it may be able to detect plasma effects that are not observable until now.

In the two examples mentioned above the gravitational field is weak, so the linearised
version of general relativity is sufficient. However, one may ask if there are effects
of a plasma on light near black holes and other compact objects where the linearised
theory is not applicable. Then the theory of general relativity has to be used without
approximations. The exact formula for the bending angle in a  
plasma whose density depends only on the radial coordinate was calculated on the
Schwarzschild spacetime and in the equatorial plane of the Kerr spacetime by
Perlick~\cite{Perlick2000b}. For further discussions of plasma effects in strong
gravitational fields see Bisnovatyi-Kogan and 
Tsupko~\cite{BisnovatyiTsupko2009,BisnovatyiTsupko2010,TsupkoBisnovatyi2013},
Morozova, Ahmedov and Tursunov~\cite{MorozovaAhmedovTursunov2013}, 
Er and Mao~\cite{ErMao2014},
Rogers~\cite{Rogers2015,Rogers2017a,Rogers2017b} and Perlick, Tsupko and 
Bisnovatyi-Kogan~\cite{PerlickTsupkoBisnovatyi2015,PerlickTsupko2017}.

Here we want to discuss the influence of a plasma on the geometry of light \emph{bundles}. 
This is of relevance for image distortion (which is coded in the shear of the bundle) 
and for distance measures (which are coded in the expansion of the bundle). 
To the best of our knowledge, this question has not been investigated
before. (There is a paper by Noonan \cite{Noonan1983} who addresses the influence
of a refractive medium on the geometry of light bundles in general relativity. Note,
however, that his equations are not valid if the index of refraction depends 
on the frequency, i.e., they do not apply to a dispersive medium such as a plasma.)
We restrict to ray optics in a non-magnetised pressurefree (cold) plasma. The fundamental 
equations, which will be briefly reviewed below, are derived and discussed 
in Perlick~\cite{Perlick2000b}.  For a magnetised plasma see Breuer and 
Ehlers~\cite{BreuerEhlers1980, BreuerEhlers1981}. For early work on light
propagation in dispersive media on a general-relativistic spacetime we refer to 
Synge~\cite{Synge1960}, Madore~\cite{Madore1974},  
Bi{\v{c}}{\'{a}}k and Hadrava~\cite{BicakHadrava1975}, 
Anile and Pantano~\cite{AnilePantano1977,AnilePantano1979}.

The paper is organised as follows: In Section~\ref{sec:vacuum} we review the 
known theory of light bundles in vacuum on a general-relativistic spacetime. 
In Section \ref{sec:plasma} we introduce the relevant definitions for light bundles
in a plasma, following the vacuum case as closely as possible; based on these
definitions, we derive the propagation equations (Sachs equations) for such 
bundles and we prove a plasma version of the reciprocity relation (Etherington law).
We illustrate the general results with a homogeneous plasma on a Robertson-Walker
spacetime in Section \ref{subsec:example} and conclude in Section \ref{sec:clonclusions}. 

We use greek indices taking values 0, 1, 2, 3 and, occasionally,  latin indices
taking values 1, 2. Greek indices, for which Einstein's summation convention is in force,
are lowered and raised with the spacetime metric $g_{\mu \nu}$ and its inverse $g^{\mu \nu}$,
respectively. We use units making $c$ and $\hbar$ equal to 1 so that there is no difference
between the 4-momentum of a light ray (i.e., of a photon) and the corresponding wave covector. 
Our choice of signature is $(-,+,+,+)$. When we use index notation we define the 
components of the Riemann tensor by $R_{\mu \nu \sigma \tau} = g \big(  \partial _{\mu} ,
\nabla _{\partial _{\nu}} \nabla _{\partial _{\sigma}} \partial _{\tau} - 
 \nabla _{\partial _{\sigma}} \nabla _{\partial _{\nu}} \partial _{\tau} \big)$ and the
components of the Ricci tensor by $R_{\nu \tau} = R^{\sigma}{}_{\nu \sigma \tau}$.

\vspace{2cm}

\section{Light propagation in vacuum}\label{sec:vacuum}
\subsection{The Hamiltonian for light rays in vacuum}
Light rays are the solutions to Hamilton's equations 
\begin{eqnarray}
\label{H-Gleichungen}
	 \frac{d x ^\mu}{ds} = \frac{\partial H}{\partial p_\mu} \, ,  \qquad 
            \frac{d p_\mu}{ds} 
	= - \frac{\partial H}{\partial x^\mu} \, ,  \qquad H(x, p) = 0 \, .
\end{eqnarray} 
Here $x = x(s)$ denotes a curve, parametrised by a real quantity $s$,  in spacetime, which describes 
the light ray and $p=p(s)$ is its canonical momentum.
For light rays in vacuum on a general-relativistic spacetime, it is well known that the Hamiltonian is 
\begin{eqnarray}
\label{H im qf Vakuum}
	H(x, p) = \frac{1}{2} g^{\mu\nu}(x)p_\mu p_\nu  \, .
\end{eqnarray}
In this case the solutions to (\ref{H-Gleichungen}) are the lightlike geodesics of the spacetime
metric. For a derivation of general-relativistic ray optics in vacuum from Maxwell's equations see,
e.g., Schneider, Ehlers and Falco~\cite{SchneiderEhlersFalco1992}. 

For the tangent vector $K^{\mu} = dx^{\mu}/ds$ of a light ray, (\ref{H im qf Vakuum})
implies
\begin{eqnarray}
\label{H-Fkt Licht Vak}
	K^\mu = \frac{\partial H}{\partial p_\mu} 
	= g^{\mu\nu} p_\nu \,,  \quad g_{\mu\nu}K^\mu 
	= 	p_\nu \, .
\end{eqnarray}
With respect to an observer with 4-velocity $U^{\nu}$, normalised as usual by the
condition $g_{\mu \nu} U^{\mu} U^{\nu} = - 1$, the tangent vector to the light ray
can be decomposed into a part parallel and  a part perpendicular to the 4-velocity
of the observer, 
\begin{eqnarray}\label{eq:komega}
K^{\mu} = \omega U^{\mu} + k^{\mu} \, , \quad U^{\nu}k_{\nu} = 0\, ,
\end{eqnarray}
where
\begin{eqnarray}\label{eq:omega}
	\omega = - p_\nu U^\nu 
\end{eqnarray}
is to be interpreted as the frequency as measured by the observer and $k^{\mu}$ is 
the spatial wave vector. This decomposition puts the dispersion relation $H(x,p)=0$ 
into the familiar form 
\begin{eqnarray}\label{eq:drvac}
\omega  = \sqrt{k_{\mu}k^{\mu}} \, .
\end{eqnarray}
The definition of the frequency 
immediately implies the redshift formula
\begin{eqnarray}\label{eq:redshift}
	\frac{\omega_{ e}}{\omega_{ r}} 
	= \frac{p_\nu U^\nu \big\vert_e}{p_\sigma U^\sigma\big\vert_r} =
           \frac{g_{\mu\nu}K^\mu U^\nu\big\vert_e}{g_{\rho\sigma}K^\rho U^\sigma\big\vert_r} 
\end{eqnarray} 
where  $e$ and $r$ stand for  ``emitter'' and ``receiver'', respectively. Owing to (\ref{eq:drvac}),
we may rewrite the redshift formula in terms of the 
wavelength $\lambda = 2 \pi / \sqrt{k_{\mu}k^{\mu}}$ rather than in terms of the frequency $\omega$, 
\begin{eqnarray}\label{eq:redshiftvac}
	\frac{\lambda_{ r}}{\lambda _{e}} = 
           \frac{\omega_{ e}}{\omega_{ r}}  
	\, . 
\end{eqnarray} 

It is often convenient to think of light propagation in terms of classical 
point particles (``classical photons'') that move along the trajectories determined by 
Hamilton's equations (\ref{H-Gleichungen}). In this interpretation we have to identify 
the frequency $\omega$ with the energy of a classical photon. In our units, making $c$ 
and $\hbar$ equal to unity, the unit of a frequency is the same as the unit of an energy, 
namely 1/length. This choice of a unit for $\omega$ fixes the unit for any other 
\emph{scalar} quantity: The Hamiltonian $H$, which should \emph{not} be confused with 
the energy of a photon, has dimension 1/length$^2$, the curve parameter $s$ has dimension
length$^2$ and $x^{\mu}p_{\mu}$ is dimensionless. The dimension of tensor components, such as
e.g. $g_{\mu \nu}$, $p_{\mu}$ or $U_{\mu}$, depend of course on the dimensions of the chosen
coordinates.

If we multiply the Hamiltonian with an arbitrary function $f(x,p)$ that has no zeros,
the light rays remain unchanged up to a reparametrisation. This will be of crucial relevance
for our treatment in the plasma case. In vacuum, however, we will always use the 
parametrisation adapted to the Hamiltonian (\ref{H im qf Vakuum}). For each light ray, 
this parametrisation is unique up to an affine reparametrisation. As usual, we refer
to it as to an \emph{affine parametrisation}. A change of the affine parametrisation
corresponds to a multiplication of $K^{\mu}$, $p_{\mu}$, $\omega$ and $k^{\mu}$
by a factor which is constant along the light ray. Note that the frequency \emph{ratio} 
(\ref{eq:redshift}) is unaffected by such a reparametrisation. 

\subsection{Light bundles in vacuum}
\label{Jacobi-Gleichung Kap}
It is now our goal to re-derive the equations that determine the shape and the 
size of light bundles in vacuum. This is known material, the relevant results go back to 
Jordan, Ehlers and Sachs~\cite{JordanEhlersSachs1961} and to Sachs~\cite{Sachs1961}.
In Section~\ref{sec:plasma} below we will then investigate if and how these results
can be generalised to the plasma case. 

Our notation follows Perlick~\cite{Perlick2004}.  For most parts of the discussion
of light bundles we find it convenient to use coordinate-free notation. Then
the tangent vector field $K$ to an affinely parametrised light ray in vacuum  has to satisfy the 
conditions $g(K,K)=0$ and  $\nabla _K K =0$, where $g$ is the spacetime metric and $\nabla$
is the Levi-Civita derivative of $g$.  

\begin{definition}[Light bundle]
\label{DefLichtbV}
Let $\lambda$ be an affinely parametrised  light ray in vacuum and $K$ its tangent vector field.
An (infinitesimally thin) \emph{light bundle} along $\lambda$ is a
set  $\mathcal{B} = \big\{ c_1 Y_1 + c_2 Y_2 \, \big|  \,  c_1^2 + c_2^2 \leq 1 \big\}$
where $Y_1$ and $Y_2$ are two vector fields along $\lambda$ such that  
\begin{itemize}
\item[\emph{(a)}] $ \nabla_K \nabla_K Y_i = R(K, Y_i,K)$ \quad for $\:  i = 1, 2$;
\item[\emph{(b)}]
	$ g(K, Y_i) = 0$  \quad for $\: i = 1, 2$;
\item[\emph{(c)}]
	$ Y_1, Y_2$ and $K$ are linearly independent almost everywhere.
\end{itemize}
\end{definition}
In condition (a) $R$ denotes the Riemannian curvature tensor. This condition is the \emph{Jacobi equation} 
(also known as the \emph{geodesic deviation equation}) which expresses the fact that $Y_1$ and $Y_2$ are
connecting vectors from the central geodesic $\lambda$ to an infinitesimally close neighbouring geodesic.
Condition (b) makes sure that these neighbouring geodesics are again lightlike and that $Y_1$ and $Y_2$ span
a spacelike plane at all points where they are linearly independent. By condition (c), this is true almost 
everywhere. This condition guarantees that the cross-section of the bundle is two-dimensional
except, possibly, at isolated points where this cross-section may collapse to a point or to a line. Such points
are known as the \emph{caustic} points of the  bundle.

As one wants to write the Jacobi equation for $Y_1$ and $Y_2$ in matrix form, it is usual to introduce an appropriate 
basis of vector fields along $\lambda$.
\begin{definition}[Sachs basis]\label{def:Sb}
A pair of vector fields, $(E_1, E_2)$, along a light ray $\lambda$ with tangent vector field $K$ 
is called a \emph{Sachs basis} if
 \begin{itemize}
	\item[\emph{(a)}]
	$ g(E_i, E_j)=\delta_{ij}$  \quad for $\:  i,j = 1, 2$ , 
	\item[\emph{(b)}]
	$ g(K, E_i) = 0$ \quad for $\: i = 1, 2$ , 
	\item[\emph{(c)}]
	$ \nabla_K E_i=0$ \quad for $\: i = 1, 2$.
\end{itemize}
\end{definition}
A Sachs basis along $\lambda$ is unique up to transformations
\begin{eqnarray}\label{eq:Etrafo}
E_1  \mapsto  \cos \alpha \, E_1 + \sin \alpha \, E_2 + a_1 K,
   \\
E_2  \mapsto  - \sin \alpha \, E_1 + \cos \alpha \, E_2 + a_2 K,
\end{eqnarray}
where $\alpha$, $a_1$ and $a_2$ are constant.
The two vectors of a Sachs basis may be interpreted as spanning a \emph{screen}. 
Choosing, at one point of $\lambda$, a timelike vector $U$ that is to be interpreted as the 4-velocity
of an observer singles out all Sachs bases perpendicular to $U$. They are unique up to spatial 
rotations, i.e., up to transformations (\ref{eq:Etrafo}) with $a_1=a_2=0$.
In this sense, the transformation (\ref{eq:Etrafo}) may be interpreted
as the combination of a spatial rotation, determined by $\alpha$, and a boost, determined
by $a_1$ and $a_2$.
 
By Definition~\ref{def:Sb} the vectors  $K$, $E_1$ and $E_2$ span the
orthocomplement of $K$ at each point of $\lambda$.  We can thus write our bundle vectors  
$Y_i$ as linear combinations of these vectors,
\begin{eqnarray}\label{eq:YE}
Y_i = D_{i1}   E_1 + D_{i2} E_2 + y_i K \,  .
\end{eqnarray}
We call the $2 \times 2$ matrix $\boldsymbol{D}= ( D_{ij} )$ the \emph{Jacobi matrix}.
It determines the shape of the cross section of the light bundle in 
the Sachs basis.


Plugging (\ref{eq:YE}) into the 
Jacobi equation and applying the operator $g(E_h, \cdot)$ yields the \emph{matrix Jacobi equation}
\begin{eqnarray}
\label{MJGl}
 \frac{d^2}{ds^2}  \boldsymbol{D}  = \boldsymbol{D} \, \boldsymbol{Z} \, .
\end{eqnarray}
Here $s$ is the affine parameter along the light ray, i.e., $\nabla _K f= df/ds$ for scalar 
functions $f$, and $\boldsymbol{Z}= (Z_{hj})$ is the \emph{optical tidal matrix}, defined
by 
\begin{eqnarray}\label{eq:Z}
Z_{hj} =    g\bigl(E_h, R(K, E_j, K)\bigr) \, .
\end{eqnarray}
$\boldsymbol{Z}$  is symmetric, $Z_{hj}=Z_{jh} $, because of the symmetry properties of 
the Riemann tensor. With the help of the well-known decomposition of the curvature tensor 
into Weyl tensor $C$ and Ricci tensor $\mathrm{Ric}$ (see 
e.g.~Wald~\cite{Wald1984}, p.~40, but note that our sign convention for the Ricci
tensor is different) the optical tidal matrix can be rewritten as 
\begin{eqnarray}
\label{Zij V}
	Z_{hj} =  \frac{1}{2} \mathrm{Ric} (K, K)  \delta _{hj} +  
          	g\bigl(E_h , C (K, E_j, K)\bigr) \, ,
\end{eqnarray} 
where we have used that $g(K,K)=0$.

\subsection{Sachs equations for light bundles in vacuum}
\label{Sachsvac}

With the \emph{deformation matrix} $\boldsymbol{S}$ defined by
 \begin{eqnarray}
\label{DefS}
	\frac{d}{ds} \boldsymbol{D} = \boldsymbol{D} \, \boldsymbol{S} \, ,
\end{eqnarray}
the matrix Jacobi equation (\ref{MJGl}) reads  
\begin{eqnarray}\label{eq:msachs}
	\frac{d}{ds} \boldsymbol{S} + \boldsymbol{S} \, \boldsymbol{S} = \boldsymbol{Z}  \, .
\end{eqnarray}
Here we have assumed that $\boldsymbol{D}$ has full rank, which is the case almost everywhere
(see Definition \ref{DefLichtbV}).
Decomposing $\boldsymbol{S}$ into antisymmetric, symmetric-tracefree and trace parts,
\begin{eqnarray}
\label{SinoptSkalare}
	\boldsymbol{S}=  \begin{pmatrix} 
		0 & - \Omega \\
		\Omega & 0
		\end{pmatrix} + \begin{pmatrix} 
		\sigma_1 & \sigma_2 \\
		\sigma_2 & - \sigma_1
		\end{pmatrix} + \begin{pmatrix} 
		\theta & 0 \\
		0 & \theta
		\end{pmatrix} \, ,
\end{eqnarray}
defines the  rotation $\Omega$, the shear $(\sigma _1 , \sigma _2)$, and the expansion $\theta$
of the  bundle which are usually combined into the two complex \emph{optical scalars}
\begin{eqnarray}\label{eq:optsc}
\rho= \theta + i \Omega \, , \quad \sigma= \sigma_1 + i \sigma_2 \, .  
\end{eqnarray} 
We read from (\ref{eq:msachs}) and (\ref{SinoptSkalare}) that the optical 
scalars have the same dimension as the curve parameter $s$, namely that of an area.
Then the matrix equation (\ref{eq:msachs}) gives the two complex \emph{Sachs equations}
\begin{eqnarray}
	\label{S1imVakuum}
		\frac{ d \rho}{ds}+ \rho^2+ | \sigma | ^2 
		= \frac{1}{2} \, \mathrm{Ric}(K,K) 
\end{eqnarray}
\begin{eqnarray}
		\label{S2imVakuum}
          \frac{d \sigma}{ds}+ 2\theta\sigma 
		=\frac{1}{2}  \, g\bigl(E_1+ iE_2, C(K, E_1+ iE_2, K)\bigr) \, .
\end{eqnarray}


Note that the deformation matrix $\boldsymbol{S}$ and, thus, the optical scalars characterise the \emph{change}
of the shape of the light  bundle. If the optical scalars are known, the shape is determined by solving the first-order
differential equation (\ref{DefS}) for the matrix $\boldsymbol{D}$. We may use a polar decomposition of $\boldsymbol{D}$,
\begin{eqnarray}\label{eq:DMR}
\boldsymbol{D} = \boldsymbol{R} \, \boldsymbol{M}
\end{eqnarray}
where $\boldsymbol{R}$ is orthogonal and $\boldsymbol{M}$ is symmetric and, thereupon, diagonalise the matrix $\boldsymbol{M}$
with the help of another orthogonal matrix. As the product of two orthogonal matrices is again orthogonal, this gives us a  
parametrisation of the matrix $\boldsymbol{D}$ in terms of two angles, $\psi$ and $\chi$, and two eigenvalues, $D_+$ and 
$D_-$,
\begin{eqnarray}\label{eq:Dpm}
\boldsymbol{D} = \begin{pmatrix}
           \cos \psi & - \sin \psi\\
	\sin \psi & \cos \psi
\end{pmatrix}
\begin{pmatrix}
	D_+ & 0 \\
	0 & D_-
\end{pmatrix}
\begin{pmatrix}
	\cos\chi & \sin\chi \\
	- \sin\chi & \cos\chi
\end{pmatrix} \, .
\end{eqnarray}
Under a transformation (\ref{eq:Etrafo}) of the Sachs basis, the four parameters of the Jacobi matrix change
according to $D_{\pm} \mapsto D_{\pm}$, $\chi \mapsto \chi - \alpha$, $\psi \mapsto \psi$. We see that 
the parameter $\alpha$ has the only effect of rotating the angle $\chi$ by a constant amount while the
parameters $a_1$ and $a_2$ have no influence at all.
This demonstrates the important fact that the shape and the size of the cross-section of a light bundle are
independent of the Sachs basis, i.e., of the chosen screen. This result was proven by Sachs~\cite{Sachs1961}
and we will refer to it as to \emph{Sachs's theorem}. 

In terms of $D_+$, $D_-$, $\psi$ and $\chi$, the matrix equation (\ref{DefS}) can be rewritten,
after a bit of algebra, as
\begin{eqnarray}\label{eq:dDpm}
\frac{d D_{\pm}}{ds} \, + \, i \, D_{\pm} \, \frac{d \chi}{ds} 
\, - \, i  \, D_{\mp} \, \frac{d \psi}{ds} \, = \, 
D_{\pm} \big( \overline{\rho \,} \, \pm \, \sigma \,
e^{- 2 i \chi} \big) 
\end{eqnarray}
where an overbar denotes complex conjugation.

\section{Light propagation in a plasma}\label{sec:plasma}
\subsection{The Hamiltonian for light rays in a plasma}\label{subsec:Hamplasma}
The Hamiltonian for light rays in a plasma  on a general-relativistic spacetime was rigorously derived by Breuer 
and Ehlers~\cite{BreuerEhlers1980,BreuerEhlers1981} from Maxwell's equations coupled to two charged fluids,
a positively charged fluid modelling the ions and a negatively charged fluid modelling the electrons. By linearising 
the equations of motion for the electrons about a background field  and assuming that only the electrons could 
follow a rapidly oscillating wave motion they derived a coupled system of linear wave equations for the 
perturbations of the electromagnetic field and of the electron fluid. The transition to ray optics was performed by 
 a two-scale method, sending to infinity simultaneously the frequency and the scale on which the background
fields vary. Breuer and Ehlers allowed for an arbitrary electromagnetic background field, i.e., they considered
a magnetised plasma, but they assumed the electron fluid to be pressureless. (This is often 
called a {\em cold} plasma.)
Then the  resulting equations of motion for the light rays could be put into Hamiltonian form (\ref{H-Gleichungen}), 
where the Hamiltonian was an eighth-order polynomial in the momentum coordinates. 
At a more fundamental level, the two-fluid plasma model which is the starting point for the derivation by Breuer 
and Ehlers can be derived from kinetic theory. In this case one would start out from the general-relativistic 
Boltzmann equation without a collision term (also known as Liouville equation or Vlasov equation) and derive
phenomenological two-fluid, or multi-fluid, flows from that, see e.g. \cite{ElsaesserPopel1997}.
In this derivation the equivalence principle is used, assuming that the plasma couples minimally to
the gravitational field, i. e., that there are no curvature couplings. With the simplifying assumption that 
the pressure can be neglected, the phenomenological plasma model of Breuer and Ehlers results.
Here, in contrast to the work of Breuer and Ehlers, we want to consider the considerably simpler case 
of a \emph{non-magnetised} pressureless plasma. For this case, the derivation of the Hamiltonian for
the light rays can also be found in the book by Perlick~\cite{Perlick2000b}. The Hamiltonian takes the 
simple  form
\begin{eqnarray}\label{eq:Hpl}
	H(x, p) = \frac{1}{2} \Bigl(g^{\mu\nu}(x)p_\mu p_\nu + \omega_p(x) ^2 \Bigr) \, ,
\end{eqnarray}
where $\omega _p (x)$ is a scalar function on the spacetime known as the \emph{plasma frequency}.
In the derivation $\omega _p (x)^2$ comes about as a multiple of the (background) number density function
$n_e$ of the electrons, as was already anticipated in Eq. (\ref{eq:fp}). As usual in relativity, $n_e$ is defined with 
respect to the rest system of the electron fluid, so it is an invariant scalar (i.e., independent of which 
coordinates are chosen on spacetime). Therefore, also the plasma frequency is an invariant scalar..

Note that the Hamiltonian (\ref{eq:Hpl}) depends on the density of the plasma but not on its 4-velocity, i.e.,
it is Lorentz invariant. Light rays in a medium whose Hamiltonian is of this form have been studied in 
detail in the book by Synge~\cite{Synge1960}. Apparently, Synge was not aware of the fact that 
a non-magnetised pressureless plasma is an example of such a medium. 

For $\omega _p \to 0$, the Hamiltonian (\ref{eq:Hpl}) reproduces of course the vacuum Hamiltonian
(\ref{H im qf Vakuum}). The most important difference between the plasma case and the  vacuum case
is in the fact that (\ref{H im qf Vakuum}) is homogeneous with respect to the momentum coordinates
while (\ref{eq:Hpl}) is not. This reflects the property of a plasma to be \emph{dispersive}, i.e., the
property that the path of a light ray depends not only on the spatial initial direction but also on the 
frequency. For a detailed discussion of ray optics in dispersive and non-dispersive media from a 
spacetime perspective we refer to Perlick~\cite{Perlick2000b}.      

As in vacuum, the tangent vector $K^{\mu} = dx^{\mu}/ds$ of a light ray satisfies
(\ref{H-Fkt Licht Vak}). The equation $H(x,p)=0$ is, thus, equivalent to 
\begin{eqnarray}\label{eq:timelike}
g_{\mu \nu} \frac{dx^{\mu}}{ds} \frac{dx^{\nu}}{ds} = - \omega _p^2 \, ,
\end{eqnarray}
which demonstrates that light rays in a plasma are \emph{timelike} curves. 

With the frequency defined by (\ref{eq:omega}), we read from (\ref{H-Fkt Licht Vak}) that  the redshift
formula (\ref{eq:redshift}) is valid in the plasma as well. If decomposed into frequency and spatial wave vector,
the dispersion relation $H(x,p)=0$ in a plasma reads
\begin{eqnarray}\label{eq:drpl}
\omega = \sqrt{\omega _p ^2 + k_{\mu} k^{\mu}} \, ,
\end{eqnarray}
which demonstrates that, at a spacetime point $x$, only frequencies $\omega > \omega _p (x)$
are possible. The limiting case $\omega = \omega _p$ corresponds to an observer whose
4-velocity $U^{\mu}$ is tangent to the light ray; for such an observer the spatial wave vector
$k^{\mu}$ is zero and (momentarily) ``the light ray stands still''. Note that, in terms of the 
decomposition (\ref{eq:komega}) with respect to an observer, the length element along the light 
ray as measured by this observer is
\begin{eqnarray}\label{eq:ell}
d \ell = \sqrt{k_{\mu}k^{\mu}} \, ds \, = \, \sqrt{\omega ^2-\omega _p^2} \, ds \, .
\end{eqnarray}

We may rewrite the redshift law (\ref{eq:redshift}) in terms of the wavelength $\lambda = 
2 \pi /\sqrt{k_{\mu}k^{\mu}}$, rather than in terms of the frequency. However, instead of 
the vacuum relation (\ref{eq:redshiftvac})  we have now to use the equation
\begin{eqnarray}\label{eq:redshiftpl}
	\frac{\lambda_{ r}}{\lambda _{e}} 
           = \frac{\sqrt{\omega_{ e}^2-\omega_{pe}^2}}{\sqrt{\omega_{ r}^2-\omega _{pr}^2}} 
\end{eqnarray} 
in accordance with the dispersion relation (\ref{eq:drpl}). Therefore,
in a plasma we have to distinguish between a wavelength redshift $z$ and a 
frequency redshift $Z$, 
\begin{eqnarray}\label{eq:zaZ}
z \equiv \frac{\lambda_r - \lambda_e}{\lambda_e}, \quad 
Z \equiv \frac{\omega_e - \omega_r}{\omega_r}.
\end{eqnarray}
They are related by the equation
\begin{eqnarray}\label{eq:zZ}
1 + Z(z,\omega_r) = (1 + z) 
\sqrt{1 + \frac{1}{\omega_r^2} \left(\frac{\omega_{pe}^2}{(1+z)^2} - \omega_{pr}^2\right)}.
\end{eqnarray} 
The easiest way of verifying this relation is to start from the right-hand side 
of (\ref{eq:zZ}) and to  express z by its defining equation,
given in (\ref{eq:zaZ}). Then substituting for $\lambda _r / \lambda _e$ from
(\ref{eq:redshiftpl}) gives immediately $\omega_e/\omega _r$, i.e, the left-hand
side of (31). 

We consider the frequency as the primary quantity and the wavelength as 
secondary. There are two reasons for this, one from a conceptual point of view and one 
from the point of view of a practitioner: Firstly, when light enters from one medium
into another its wavelength changes while its frequency remains the same. Secondly,  
plasma effects are non-negligible only in the radio regime where common spectroscopic 
measuring devices determine the frequency and not the wavelength. This is in agreement
with what Appenzeller \cite{Appenzeller2013} writes on p.54 of his text-book on
astronomical spectroscopy: ``In general, using the frequency is preferable as (it) 
is an intrinsic property of the radiation, whereas the wavelength depends on the medium 
in which the wave is propagating ... Counting oscillations is relatively easy in the 
radio bands up to 100 GHz ...''. On the other hand, the wavelength redshift is also of
relevance. E.g., modern arrays of radio telescopes like LOFAR, MWA, SKA, etc are 
related to wavelength measurements. Also, we will see in Section \ref{subsec:example}
below that for a homogeneous plasma on a Robertson-Walker spacetime the wavelength redshift 
is related to the scale factor by the same equation as in vacuum whereas the frequency 
redshift is given by a rather complicated expression.

We have already mentioned that, up to parametrisation, the solutions to the system 
(\ref{H-Gleichungen}) are unchanged if we multiply the Hamiltonian with a nowhere 
vanishing function. In the plasma case it is often advantageous to  
switch to the dimensionless Hamiltonian
\begin{eqnarray} \label{eq:tH}
\tilde{H} (x, p ) = \omega _p (x)^{-2} \, H(x,p) =
\frac{1}{2} \Big( \omega _p (x) ^{-2} g^{\mu \nu} (x) p_{\mu} p_{\nu} \, + \, 1 \, \Big)
\, .
\end{eqnarray}
This is possible on any domain where $\omega _p$ has no zeros. Solving Hamilton's equations
\begin{eqnarray}\label{tH-Gleichungen}
	 \frac{d x ^\mu}{d \tilde{s}} = \frac{\partial \tilde{H}}{\partial p_\mu} \, ,  \qquad 
            \frac{d p_\mu}{d \tilde{s}} 
	= - \frac{\partial \tilde{H}}{\partial x^\mu} \, ,  \qquad \tilde{H} (x, p) = 0 \, 
\end{eqnarray} 
with this transformed Hamiltonian shows that the light rays in the plasma are 
the \emph{timelike} geodesics of the conformally rescaled  metric
\begin{eqnarray}\label{eq:tg}
\tilde{g}{}_{\mu \nu} (x) = \omega _p (x) ^2 \, g_{\mu \nu} (x) \, .
\end{eqnarray}
Up to a constant factor, the dimensionless new parameter $\tilde{s}$ is proper time with 
respect to the metric $\tilde{g}{}_{\mu \nu}$. It is related to the old parameter $s$ by
\begin{eqnarray}\label{eq:sts}
\frac{d}{ds} = \omega _p^{-2} \, \frac{d}{d \tilde{s}}\, , 
\end{eqnarray}
in particular
\begin{eqnarray}\label{eq:tK}
K^{\mu} = \omega _p^{-2} \widetilde{K}{}^{\mu} \, , \quad
K^{\mu} = \frac{dx^{\mu}}{d s}  \, , \quad 
\widetilde{K}{}^{\mu} = \frac{d x ^{\mu}}{d \tilde{s}} \, .
\end{eqnarray}
With the help of (\ref{eq:tK}) one may rewrite the redshift law  (\ref{eq:redshift}) 
in terms of $\widetilde{K}{}^{\mu}$,
\begin{eqnarray}\label{eq:tredshift}
	\frac{\omega_{ e}}{\omega_{ r}} =
           \frac{
           \tilde{g}{}_{\mu\nu} \widetilde{K}{} ^\mu U^\nu\big\vert_e
           }{
           \tilde{g}{}_{\rho\sigma}\widetilde{K}{}^\rho U^\sigma\big\vert_r
           }  \, .
\end{eqnarray}

\subsection{Light bundles in a plasma}
\label{subsec:bundlepl}
For the following considerations we restrict ourselves to a spacetime region where the
plasma density $\omega _p (x)$  has no zeros. On this region the conformally
rescaled metric (\ref{eq:tg}) is well defined and we may use for light rays the 
parametrisation adapted to the Hamiltonian $\tilde{H}$. Then the tangent vector field
$\widetilde{K}$ to a light ray has to satisfy the equations $\tilde{g}\big( \widetilde{K},
\widetilde{K} \big) = - 1$  and $\widetilde{\nabla} 
_{\widetilde{K}} \widetilde{K} = 0$ where $\widetilde{\nabla}$ is the Levi-Civita
derivative of the metric $\tilde{g}$.

We now define the notion of a light bundle in the plasma, following the construction
for the vacuum case as closely as possible.  

\begin{definition}[Light bundle]
\label{DefLichtbP}
Let $\tilde{\lambda}$ be a  light ray in the plasma, parametrised by proper time 
$\tilde{s}$ with respect to the
metric $\tilde{g}$, and $\widetilde{K}$ its tangent vector field.
An (infinitesimally thin) \emph{light bundle} along $\tilde{\lambda}$ is a
set  $\mathcal{B} = \big\{ c_1 \widetilde{Y}{}_1 + c_2 \widetilde{Y}{}_2 \, \big|  \,  c_1^2 + c_2^2 \leq 1 \big\}$
where $\widetilde{Y}{}_1$ and $\widetilde{Y}{}_2$ are two vector fields along $\tilde{\lambda}$ such that  
\begin{itemize}
\item[\emph{(a)}] $ \widetilde{\nabla}_{\widetilde{K}} \widetilde{\nabla}_{\widetilde{K}} \widetilde{Y}_i 
                             = \widetilde{R}(\widetilde{K}, \widetilde{Y}_i, \widetilde{K})$ \quad for $\:  i = 1, 2 \,$;
\item[\emph{(b)}]
	$ \tilde{g}(\widetilde{K}, \widetilde{Y}_i) = 0$  \quad for $\:  i = 1, 2 \,$;
\item[\emph{(c)}]
	$ \widetilde{Y}_1$ and $\widetilde{Y}_2$ are linearly independent almost everywhere.
\end{itemize}
\end{definition}
In condition (a), $\tilde{R}$ denotes the Riemannian curvature
tensor of the metric $\tilde{g}$. This condition makes sure that $\widetilde{Y}{}_1$ 
and $\widetilde{Y}{}_2$ are connecting vectors
to neighbouring $\tilde{g}-$geodesics. Condition (b) implies that these neighbouring geodesics are again
timelike and parametrised by proper time with respect to $\tilde{g}$, and that  $\widetilde{Y}{}_1$ and 
$\widetilde{Y}{}_2$ 
span a \emph{spacelike} plane at each point where they are linearly independent. The third 
condition makes sure that this is true almost everywhere, i.e., that the bundle has a 
two-dimensional cross-section except, possibly, at some caustic points.

Here it is important to realise that Definition~\ref{DefLichtbP} allows the neighbouring light rays to have 
arbitrary frequencies. We do not attempt to define something like ``a bundle of light rays with a given
frequency'', and it is hard to see how this could be done. Even in cases where we have a distinguished
observer field with respect to which the frequency could be defined, we would have to face the problem
that the frequency is not in general conserved along a light ray. 

The following definition is crucial for all that follows.
\begin{definition}[Sachs basis]
\label{def:Sachspl}
A pair of vector fields, $( \widetilde{E}{}_1, \widetilde{E}{}_2)$, along a light ray $\tilde{\lambda}$ with tangent vector field 
$\widetilde{K}$, is called a \emph{Sachs basis}  if
 \begin{itemize}
	\item[\emph{(a)}]
	$ \tilde{g}(\widetilde{E}_i, \widetilde{E}_j)=\delta_{ij}$ \quad for $\: i,j = 1, 2$,
	\item[\emph{(b)}]
	$ \tilde{g}(\widetilde{K}, \widetilde{E}_i) = 0$ \quad for $\: i = 1, 2$,
	\item[\emph{(c)}]
	$ \widetilde{\nabla}_{\widetilde{K}} \widetilde{E}_i=0$ \quad for $\: i = 1, 2$.
\end{itemize}
\end{definition}
The choice of a timelike vector $U$ at one point of $\tilde{\lambda}$ that is not tangent to $\tilde{\lambda}$ 
singles out those Sachs bases that are perpendicular to $U$. As in the vacuum case, they are unique up to
transformations
\begin{eqnarray}\label{eq:Erot}
\widetilde{E}{}_1 \mapsto  \cos \alpha \, \widetilde{E}{}_1 + \sin \alpha \, \widetilde{E}{}_1
\\
\widetilde{E}{}_2 \mapsto - \sin \alpha \, \widetilde{E}{}_1 + \cos \alpha \, \widetilde{E}{}_2
\end{eqnarray}
with a \emph{constant} $\alpha$. However, two  \emph{arbitrary} Sachs bases along a light ray in the plasma
are related by a transformation that is quite different from the vacuum case (\ref{eq:Etrafo}). 
We will not write down such a general transformation which involves a rotation in all three
spatial dimensions.

In the vacuum case, the three vectors $K$, $E_1$ and $E_2$ spanned the orthocomplement of
$K$ which was lightlike; as the vector fields $\widetilde{Y}{}_1$ and 
$\widetilde{Y}{}_2$ were in this orthocomplement, they could be written as
a linear combination of these three vectors. By contrast, in a plasma the 
vectors $\widetilde{K}$, $\widetilde{E}{}_1$ and $\widetilde{E}{}_2$ 
span a 3-dimensional \emph{timelike} hyperplane in the tangent space
at each point, so $\widetilde{Y}{}_1$ and $\widetilde{Y}{}_2$ cannot be
written as a linear combination of them. However, we may write 
these two vector fields in the form
\begin{eqnarray}
	\label{DE}
	\widetilde{Y}_i = \widetilde{D}{}_{i1} \widetilde{E}_1 + 
           \widetilde{D}{}_{i2} \widetilde{E}_2\,+ \widetilde{Y}{}_i ^{\perp}
\end{eqnarray}
where $ \widetilde{Y}{}_i ^{\perp}$ is orthogonal to  $\widetilde{K}$, $\widetilde{E}{}_1$ and
$\widetilde{E}{}_2$. If we interpret $\widetilde{E}_1$ and $\widetilde{E}_2$ as two 
vectors that span a screen, the component $\widetilde{Y}{}_i ^{\perp}$ is
perpendicular to the screen and, thus, irrelevant for the shape of the bundle 
on the screen. Hence, the shape of the bundle 
on the screen is determined by the matrix $\widetilde{\boldsymbol{D}} = 
\big( \widetilde{D}{}_{ij} \big)$. The bundle cross-section on the screen is 
2-dimensional if the matrix $\tilde{\boldsymbol{D}}$ is non-degenerate. By condition
(c) of Definition~\ref{DefLichtbP}, this is true at almost all points of the bundle for 
almost all choices of the Sachs basis. (However, in contrast to the vacuum case, in the
plasma it is possible to make a ``bad choice'' of the Sachs basis such that the screen
is not transverse to the bundle.)  

Plugging (\ref{DE}) into the Jacobi equation for the $\widetilde{Y}_i$ and applying 
the operator $\tilde{g}(\tilde{E_h}, \cdot)$ leads to the matrix Jacobi equation in 
the plasma,
\begin{eqnarray}
\label{MJGlPl}
	 \frac{d^2}{d \tilde{s}{}^2}  \widetilde{\boldsymbol{D}}
	= \widetilde{ \boldsymbol{D}} \widetilde{\boldsymbol{Z}}
\end{eqnarray}
where $\tilde{s}$ is the parameter along the light ray (i.e., proper time with
respect to the metric $\tilde{g}$) and 
$\widetilde{\boldsymbol{Z}}= (\widetilde{Z}{}_{jh})$ with 
\begin{eqnarray}
	\widetilde{Z}_{hj}
	= \tilde{g} (\widetilde{E}_h, \widetilde{R} ( \widetilde{K}, \widetilde{E}_j, \widetilde{K})) \, .
\end{eqnarray}
In analogy to the vacuum case, we may use the symmetry of the Riemann tensor $\widetilde{R}$
to conclude that
\begin{eqnarray}\label{eq:tZsym}
           \widetilde{Z}_{hj} = \widetilde{Z}_{jh} \, ,
\end{eqnarray}
and we may decompose $\widetilde{R}$ into the Ricci tensor $\widetilde{\mathrm{Ric}}$,
the Ricci scalar $\tilde{\mathcal{R}}$ 
and the Weyl tensor $\widetilde{C}$, 
\begin{eqnarray}\label{eq:ZC}
\widetilde{Z}_{hj}
	=  \frac{1}{2} \widetilde{\mathrm{Ric}} ( \widetilde{K}, \widetilde{K}) \delta_{hj}
             + \tilde{g} (\widetilde{E}_h, \widetilde{C} ( \widetilde{K}, \widetilde{E}_j, \widetilde{K}))
\\
\nonumber
	+ \tilde{g} ( \widetilde{K}, \widetilde{K}) \Big( \frac{1}{2} \widetilde{\mathrm{Ric}} ( \widetilde{E}_h, \widetilde{E}_j)  
	- \frac{1}{3} \, \widetilde{\mathcal{R}} \, \delta_{hj} \Big)  
\end{eqnarray}
where $\tilde{g} ( \widetilde{K} , \widetilde{K} ) = -1$.

\subsection{Sachs equations for light bundles in a plasma}
\label{Sachspl}
In analogy to the vacuum case, we may introduce the
deformation matrix $\widetilde{\boldsymbol{S}}$ in the plasma  by
 \begin{eqnarray}
\label{Def tildeS}
	\frac{d}{d \tilde{s}} \widetilde{\boldsymbol{D}} = 
          \widetilde{\boldsymbol{D}}\widetilde{\boldsymbol{S}} \, .
\end{eqnarray}
Then the matrix Jacobi equation (\ref{MJGlPl}) implies
\begin{eqnarray}\label{eq:msachst}
\label{DG Stilde}
 \frac{d}{d \tilde{s}}  \widetilde{\boldsymbol{S}} + 
\widetilde{\boldsymbol{S}}\widetilde{\boldsymbol{S}} = \widetilde{\boldsymbol{Z}} 
\, .
\end{eqnarray}
From decomposing $\widetilde{\boldsymbol{S}}$ into antisymmetric, symmetric-tracefree
and trace parts we would get a set of Sachs equations in terms of the twiddled quantities. 
However, in view of applications to physics this is not the kind of Sachs equations we want 
to have in a plasma. The reason is that, according to item (a) of Definition~\ref{def:Sachspl}, 
our Sachs basis vectors are normalised with respect to the metric $\tilde{g}$. As a 
consequence, the matrix $\widetilde{\boldsymbol{D}}$ gives us the bundle cross-section
as it is measured with the metric $\tilde{g}$. Actually, in view of applications to physics
we need the bundle cross-section as it is measured with the spacetime metric $g$. 
Therefore, we introduce a rescaled Sachs basis,
\begin{eqnarray}\label{eq:Epl}
	E_i = \omega_p \widetilde{E}_i \, ,
\end{eqnarray}
and a correspondingly rescaled Jacobi matrix
\begin{eqnarray}\label{eq:DtD}
\boldsymbol{D} = \omega _p ^{-1} \, \widetilde{\boldsymbol{D}} \, .
\end{eqnarray}
The physically relevant optical scalars have to be defined in terms of this 
modified Jacobi matrix. At the same time, we may change from the parametrisation
by $\tilde{s}$ to the parametrisation by $s$ which is given by (\ref{eq:sts}). This
has the advantage that the parameter $s$, in contrast to the parameter $\tilde{s}$,
 remains meaningful in the vacuum limit $\omega _p \to 0$. 

It is now straight forward, though somewhat tedious, to derive the Sachs 
equations in a plasma. On the left-hand side of  (\ref{eq:msachst}), 
we rewrite $d/d \tilde{s}$ in terms of $d/ds$ with the help of (\ref{eq:sts}), 
and we express the matrix $\tilde{\boldsymbol{S}}$
in terms of the matrix $\boldsymbol{S}$ defined by
\begin{eqnarray}\label{eq:Spl}
	\frac{d}{ds} \boldsymbol{D} = \boldsymbol{D} \, \boldsymbol{S}
\, ,
\end{eqnarray}
hence
\begin{eqnarray}\label{eq:StS}
\boldsymbol{S} = \omega_p ^{-2} \, \widetilde{\boldsymbol{S}}
- \omega _p ^{-1} \, \frac{d \omega _p}{ds} \, \boldsymbol{1} \, .
\end{eqnarray} 
On the right-hand side of  (\ref{eq:msachst}), we express the Ricci tensor and the Weyl tensor of the 
metric $\tilde{g}$ in  terms of the Ricci tensor and the Weyl tensor of the 
conformally related metric $g$, using the well-known transformation formulas 
which are given,  e.g., in Appendix D of the book by Wald~\cite{Wald1984}. 
If we define the optical scalars as in vacuum by (\ref{SinoptSkalare}) and
(\ref{eq:optsc}), we get the following set of Sachs equations:
\begin{eqnarray}\label{S_IPlasma}
\fl \quad 		\frac{d \rho}{ds}+ \rho^2+ | \sigma |^2
		 = \frac{1}{2} R_{\mu\nu} K^\mu K^\nu
			+ \frac{1}{2} C _{\mu\nu\sigma\rho}  
                                   K^\nu K^\rho
                                   \bigl( E_1^\mu E_1^\sigma  +  E_2^\mu E_2^\sigma \bigr)
\\
\nonumber
	 + 2(\nabla^\nu \omega_p) (\nabla_\nu \omega_p)+ \omega_p (\nabla^\nu \nabla_\nu \omega_p)
 		 - \frac{1}{4}\omega_p^2 R_{\mu \nu} \Bigl( E_1^\mu E_1^\nu +   E_2^\mu E_2^\nu\Bigr)
\\
\nonumber
		  + \Big( (\nabla_\mu \omega_p)(\nabla_\nu \omega_p )- \frac{1}{2} \omega_p (\nabla_\mu \nabla_\nu \omega_p ) \Big) 
                      \Big( E_1^\mu E_1^\nu + E_2^\mu E_2^\nu \Big) 
		 + \frac{1}{3} \, \omega_p^2 \, \mathcal{R} \, , 
\end{eqnarray}
\begin{eqnarray}\label{S_IIPlasma}
\fl \quad		\frac{ d \sigma}{ds}+2\theta\sigma
		= \frac{1}{2} C _{\mu\nu\sigma\rho}  K^\nu  K^\rho
                     \Big( E_1^\mu+i E_2 ^{\mu} \Big) \Big(   E_1^\sigma  + i  E_2^\sigma  \Big)
\\
\nonumber		+ \Big( (\nabla_\mu \omega_p)(\nabla_\nu \omega_p)  -
 \frac{1}{2}\omega_p (\nabla_\mu \nabla_\nu \omega_p ) 
                      - \frac{1}{4} R_{\mu\nu}  \omega_p^2 \Big)
	           (E^\mu_1 E^\nu_1 - E^\mu_2 E^\nu_2) 
\\
\nonumber
		 - i \Big( \frac{1}{2} \omega_p^2 R_{\mu\nu} + \omega_p (\nabla_\mu \nabla_\nu \omega_p ) - 
                      2 (\nabla_\mu \omega_p) (\nabla_\nu \omega_p ) \Big) E_1^\mu E_2^\nu 
                      \, . 
	\end{eqnarray}
Note that the Weyl tensor term in (\ref{S_IPlasma}) goes to 0 if $K$ becomes lightlike. Hence, for  $\omega_p \to 0$ 
these equations reduce indeed to the Sachs equations in vacuum.

As the relation between the Jacobi matrix $\boldsymbol{D}$ and the deformation matrix $\boldsymbol{S}$
is unchanged in comparison to the vacuum case, equation (\ref{eq:dDpm}) is still valid in the plasma if we
parametrise the Jacobi matrix according to (\ref{eq:Dpm}). However, in the plasma there is no analogue to
Sachs's theorem: The eigenvalues $D_+$ and $D_-$ of the Jacobi matrix and, thus, the shape and the size
of a bundle cross-section depend on the Sachs basis. For a given Sachs basis, the corresponding values
$D_+$ and $D_-$ give the size of the cross-section as measured by an observer only for those observers
whose 4-velocity is orthogonal to the Sachs basis vectors.

\subsection{Reciprocity theorem and  related results}
\label{SvS V}

In vacuum it is well known that for Jacobi matrices a certain conservation law is satisfied. This gives
rise to the so-called reciprocity theorem (also known as the Etherington law \cite{Etherington1933})
which is of great relevance in view of physics; among other
things,  it relates the area distance to the luminosity distance which is particularly important for cosmology.
In this section we will prove a generalisation of this conservation law for light bundles in a plasma.
For a detailed discussion of the vacuum case see, e.g., Perlick~\cite{Perlick2004}.

When we defined  light bundles in a plasma we considered a central light ray with a parameter
$\tilde{s}$  adapted to the Hamiltonian $\tilde{H}$ of (\ref{eq:tH}). We then switched, by
(\ref{eq:sts}), to a parameter $s$ adapted to the Hamiltonian $H$ of (\ref{eq:Hpl}). 
The parametrisation with respect to $s$ has the obvious advantage that it remains meaningful in the
limit $\omega _p \to 0$ where the light rays become lightlike geodesics of the spacetime metric and 
$s$ becomes an affine parameter. For this reason, we use the parametrisation by $s$ in the following
theorem.
\begin{theorem}
Let $\lambda$ be a light ray in a plasma, parametrised by the parameter $s$ adapted to the 
Hamiltonian \emph{(\ref{eq:Hpl})}. Let $\boldsymbol{D}{}_1$ and $\boldsymbol{D}{}_2$ be the 
Jacobi matrices \emph{(\ref{eq:DtD})} associated with two bundles along this light ray. Then 
\begin{eqnarray}
	\frac{d}{ds} \bigg( \Big( \frac{d}{d s} \boldsymbol{D}{}_1\Big) \boldsymbol{D}{}_2^T - 
           \boldsymbol{D}{}_1 \Big(\frac{d}{d s} \boldsymbol{D}{}_2^T \Big) \bigg) = \boldsymbol{0} 
\end{eqnarray}
where $( \, . \, ) ^T$ means the transpose of a matrix.
\end{theorem}

\noindent
{\bf Proof}: By (\ref{MJGlPl}), 
\begin{eqnarray}
\fl \qquad 
\frac{d}{d \tilde{s}}\bigg( \Bigl(\frac{d}{d \tilde{s}} 
		\widetilde{\boldsymbol{D}}{}_1 \Big) \widetilde{\boldsymbol{D}}{}_2^T 
		- \widetilde{\boldsymbol{D}}{}_1 \Big(\frac{d}{d \tilde{s}} \widetilde{\boldsymbol{D}}{}_2^T\Big)\bigg) 
\\
\nonumber
\fl \qquad		= \Big(\frac{d^2}{d \tilde{s}{}^2} \widetilde{\boldsymbol{D}}{}_1\Big)
 		\widetilde{\boldsymbol{D}}_2^T + \Big(\frac{d}{d \tilde{s}} \widetilde{\boldsymbol{D}}{}_1\Big) 
                      \Big(\frac{d}{d \tilde{s}}\widetilde{\boldsymbol{D}}_2^T \Big)
 		- \Big(\frac{d}{d \tilde{s}} \widetilde{\boldsymbol{D}}{}_1\Big) 
                      \Big(\frac{d}{d \tilde{s}} \widetilde{\boldsymbol{D}}_2^T\bigr)
 		- \widetilde{\boldsymbol{D}}_1 \bigl(\frac{d^2}{d \tilde{s}{}^2} \widetilde{\boldsymbol{D}}_2^T\bigr) 
\\
\nonumber
\fl \qquad
		= \widetilde{\boldsymbol{D}}{}_1 \widetilde{\boldsymbol{Z}} \widetilde{\boldsymbol{D}}_2^T - 
                      \widetilde{\boldsymbol{D}}_1 (\widetilde{\boldsymbol{D}}_2 \widetilde{\boldsymbol{Z}})^T 
		= \widetilde{\boldsymbol{D}}_1 \widetilde{\boldsymbol{Z}} \widetilde{\boldsymbol{D}}_2^T - 
                     \widetilde{\boldsymbol{D}}_1 \widetilde{\boldsymbol{Z}}^T \widetilde{\boldsymbol{D}}_2^T 
                      = \boldsymbol{0} \, ,
\end{eqnarray}
where we have used that $\widetilde{\boldsymbol{Z}}^T=\widetilde{\boldsymbol{Z}}$. Hence
\begin{eqnarray}
\fl \qquad
	\boldsymbol{0} = \frac{d}{ds} \bigg(  
		\Big(\frac{d}{d \tilde{s}} \widetilde{\boldsymbol{D}}_1\Big) \widetilde{\boldsymbol{D}}_2^T 
		- \widetilde{\boldsymbol{D}}{}_1 \Big(\frac{d}{d \tilde{s}} \widetilde{\boldsymbol{D}}_2^T\Big) \bigg) 
\\
\nonumber
\fl \qquad
		= \frac{d}{ds} \bigg(
                      \omega _p ^{-2} \frac{d}{d s} \Big( \omega_p \boldsymbol{D}{}_1\Big) \omega_p \boldsymbol{D}_2^T 
		- \omega_p \boldsymbol{D}{}_1 \omega _p ^{-2} \frac{d}{d s}
                      \Big( \omega_p \boldsymbol{D}{}_2^T\Big) \bigg)
\\
\nonumber
\fl \qquad
		= \frac{d}{ds} \bigg(
                     \omega _p ^{-1} \frac{d \omega _p}{d s}  \boldsymbol{D}{}_1  \boldsymbol{D}{}_2^T + 
                     \Big( \frac{d }{d s} \boldsymbol{D}{}_1 \Big) \boldsymbol{D}{}_2^T 
		- \omega_p ^{-1} \boldsymbol{D}{}_1  \frac{d \omega _p}{d s} \boldsymbol{D}_2^T 
                     - \boldsymbol{D}{}_1  \Big( \frac{d }{d s} \boldsymbol{D}{}_2^T \Big)  \bigg)
\\
\nonumber
\fl \qquad
		= \frac{d}{ds} \bigg( 
                     \Big(  \frac{d}{d s} \boldsymbol{D}{}_1 \Big)  \boldsymbol{D}{}_2^T -
                     \boldsymbol{ D}{}_1 \Big(\frac{d}{d s} \boldsymbol{D}{}_2^T \Big)
                     \bigg) 
		\, .  
\end{eqnarray}
\hfill $\Box$

\noindent
The conservation law can be applied to the case that $\boldsymbol{D}{}_1 = \boldsymbol{D}{}_2=
\boldsymbol{D}$. We will now prove that this implies an important property of homocentric bundles, i.e., of 
bundles where $\boldsymbol{D}$ vanishes at a certain point.

\begin{theorem}
A homocentric bundle is twistfree.
\end{theorem}

\noindent
{\bf Proof}: For a homocentric  bundle with Jacobi matrix $\boldsymbol{D} =
\boldsymbol{D}{}_1 = \boldsymbol{D}{}_2$,  the conservation law implies that
\begin{eqnarray}\label{eq:homo}
\fl \quad                   \boldsymbol{0}  = \Big(  \frac{d}{d s} \boldsymbol{D} \Big)  \boldsymbol{D}{}^T -
                     \boldsymbol{ D}  \Big(\frac{d}{d s} \boldsymbol{D}{}^T \Big)
                     = \boldsymbol{D}  \boldsymbol{S}   \boldsymbol{D}{}^T-
                     \boldsymbol{ D}  \Big( \boldsymbol{D} \boldsymbol{S} \Big) ^T = 
                     \boldsymbol{D}  \Big( \boldsymbol{S}  - \boldsymbol{S} {}^T \Big)  \boldsymbol{D}{}^T
                     \, .
\end{eqnarray}
As $\boldsymbol{D}$ is generically invertible, this implies that $\boldsymbol{S} = \boldsymbol{S}{}^T$.
By (\ref{SinoptSkalare}), $\Omega =0$.

\hfill $\Box$

\noindent
The most important consequence of the conservation law is the following theorem.
\begin{theorem}[Reciprocity Theorem]\label{theo:reci}
Let $\lambda$ be a light ray in a plasma, parametrised by the parameter $s$ adapted to the 
Hamiltonian \emph{(\ref{eq:Hpl})}. Let $\boldsymbol{D}{}_1$ and $\boldsymbol{D}{}_2$ be 
the Jacobi matrices \emph{(\ref{eq:DtD})} associated with two bundles along this 
light ray with
\begin{eqnarray}\label{eq:in1}
	 \boldsymbol{D}{}_1 (s)\big\vert_{s = s_1} = \boldsymbol{0} \, , \quad 
	 \frac{d}{d s} \boldsymbol{D}{}_1 (s)\big\vert_{s = s_1} =  - c_1 \boldsymbol{1}  \, , 
\end{eqnarray}
\begin{eqnarray}\label{eq:in2}
 	\boldsymbol{D}{}_2 (s)\big\vert_{s = s_2}= \boldsymbol{0} \, , \quad 
	\frac{d}{d s} \boldsymbol{D}{}_2 (s)\big\vert_{s = s_2} =  c_2 \boldsymbol{1}  \, ,
\end{eqnarray}
where $c_1$ and $c_2$ are positive constants.
Then 
\begin{eqnarray}
\label{Rezi Pla}
	c_1 \boldsymbol{D}{}_2^T(s)\Big\vert_{s = s_1} =  c_2 \boldsymbol{D}_1(s)\Big\vert_{s = s_2} \; .
\end{eqnarray}
If we parametrise the Jacobi matrices according to \emph{(\ref{eq:Dpm})}, 
\begin{eqnarray}
\label{Rezi mit D+D-}
	c_1 \sqrt{|D_{2+}(s)  D_{2-}(s)|}\Big\vert_{s = s_1} 
	= c_2 \sqrt{|D_{1+} (s) D_{1-}(s)|} \Big\vert_{s = s_2} \, .
\end{eqnarray}
\end{theorem}

\noindent
{\bf Proof}: The conservation law implies that
\begin{eqnarray}\label{eq:cD1D2}
\fl \quad                      \bigg( \Big(  \frac{d}{d s} \boldsymbol{D}{}_1 \Big)  \boldsymbol{D}{}_2^T -
                     \boldsymbol{ D}{}_1 \Big(\frac{d}{d s} \boldsymbol{D}{}_2^T \Big) \bigg) \bigg|_{s=s_1}
=
                     \bigg( \Big(  \frac{d}{d s} \boldsymbol{D}{}_1 \Big)  \boldsymbol{D}{}_2^T -
                     \boldsymbol{ D}{}_1 \Big(\frac{d}{d s} \boldsymbol{D}{}_2^T \Big)\bigg)  \bigg| _{s=s_2}
\end{eqnarray}
Inserting the initial conditions (\ref{eq:in1}) and (\ref{eq:in2}) yields (\ref{Rezi Pla}).
(\ref{Rezi mit D+D-}) follows by taking the determinant.
\hfill $\Box$

\noindent
The reciprocity theorem for light bundles in vacuum was proven by  Etherington~\cite{Etherington1933}.

\begin{figure}\label{fig:reci}
\includegraphics[width=0.9\textwidth]{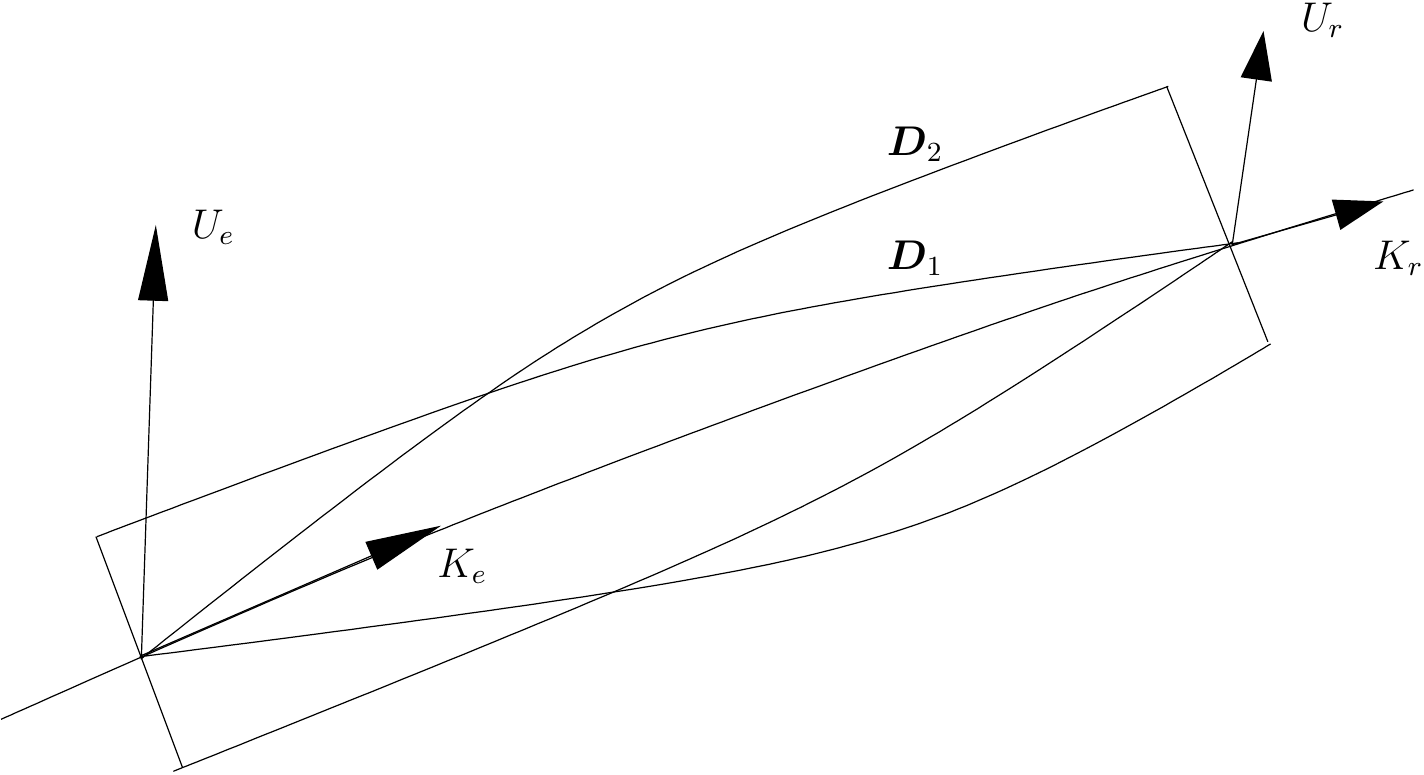}
\caption{Illustration of the reciprocity theorem.}
\end{figure}

We will now discuss the notions of area distance and luminosity distance and their relation in  a
plasma. To that end we have to consider two bundles along the same light ray, one with a vertex
at the receiver and one with a vertex at the emitter, see Fig.~\ref{fig:reci}, and we have to
apply the reciprocity theorem. For a discussion of area distance and luminosity distance in vacuum
see e.g. Schneider, Ehlers and Falco~\cite{SchneiderEhlersFalco1992} or Perlick~\cite{Perlick2004}.
 
The \emph{area distance} $D_A$ of an emitter from a receiver is defined as
\begin{eqnarray}\label{eq:DA}
D_A^2 = \frac{\text{cross-sectional  area  at  the  emitter}}{\text{solid   angle  at  the  receiver}}
\, .
\end{eqnarray}
We could directly observe the area distance if we had \emph{standard rulers}, i.e., 
light sources of a known cross-sectional area. Then the apparent size of such an object in the
sky would give us directly its area distance. While we do not have perfect standard rulers, we may at 
least estimate reasonably well the cross-sectional area of some objects (e.g. galaxies of a certain type) 
which allows us to determine the area distance of these objects to within 
a certain accuracy. (Instead of the area, one could consider all angular diameters of
the cross-section. The resulting distance measures are called the \emph{angular-diameter distances}. The
area distance is the average of the angular-diameter distances over all transverse directions. In an 
isotropic situation the notions of  area distance and angular-diameter distance coincide.) 
To work out this definition in a plasma, we consider a light ray parametrised by the parameter $s$ adapted
to the Hamiltonian (\ref{eq:Hpl}), i.e., the tangent vector $K$ of the light ray satisfies $g(K,K)=-
\omega _p^2$. The parameter runs in the future direction from the value $s_e$ at the emission event 
to the value $s_r$ at the reception event. We choose a Sachs basis along the light ray and 4-velocity 
vectors $U_e$ and $U_r$ at the emission event and at the reception event, respectively, which are 
both assumed to be orthogonal to the Sachs basis. We consider a bundle with a vertex at the receiver,
\begin{eqnarray}\label{eq:vertex}
\boldsymbol{D}{}_1 (s) \big| _{s=s_r} = \boldsymbol{0} \, , \quad
\frac{d}{ds} \boldsymbol{D}{}_1 (s) \big| _{s=s_r}  = - \omega _r \boldsymbol{1} 
\end{eqnarray}
where $\omega _r$ is the frequency of the light ray with respect to an observer with 4-velocity $U_r$. The solid 
angle of this bundle at  the receiver is assumed to be measured with respect to the same observer. The 
cross-sectional area at the emission event  is assumed to be measured by an observer with 4-velocity $U_e$.
It is crucial to keep in mind that the following calculation applies only to the case that both $U_e$ and $U_r$
are orthogonal to the chosen Sachs basis.

If the bundle is parametrised according to (\ref{eq:Dpm}), the definition of the area distance
 (\ref{eq:DA}) can be rewritten as
\begin{eqnarray}\label{eq:DA2}
D_A^2 = 
\frac{
\big| D_{1+}(s_e) D_{1-} (s_e) \big|
}{
\underset{\varepsilon \to 0}{\mathrm{lim}} 
\frac{
\big| D_{1+}(s_r-\varepsilon ) D_{1-} (s_r-  \varepsilon ) \big|
}{
\ell (\varepsilon) ^2
}
}
\end{eqnarray}
where $\ell ( \varepsilon )$ is the length of the segment of the light ray from $s_r- \varepsilon$ to $s_r$,
as measured by the observer with 4-velocity $U_r$. As we consider the limit $\varepsilon \to 0$, we need
$\ell (\varepsilon)$ only to within linear order which can be read from (\ref{eq:ell}),
\begin{eqnarray}\label{eq:ellepsilon}
\ell (\varepsilon ) = \sqrt{\omega _r^2-\omega _{pr}^2} \;  \varepsilon \, + \, \dots
\end{eqnarray}
where $\omega _{pr}$ is the plasma frequency at the reception event.
Thus, (\ref{eq:DA2}) reads
\begin{eqnarray}
D_A^2 = 
\frac{
\big| D_{1+}(s_e) D_{1-} (s_e) \big| \, \big( \omega _r ^2 - \omega _{pr}^2 \big)
}{
\underset{\varepsilon \to 0}{\mathrm{lim}} 
\frac{
\big| D_{1+}(s_r-\varepsilon ) D_{1-} (s_r-  \varepsilon ) \big|
}{
\varepsilon  ^2
}
}
\, .
\end{eqnarray}
As the vertex condition (\ref{eq:vertex}) implies that 
\begin{eqnarray}
 \frac{dD_{1+}(s)}{ds} \Big| _{s=s_r} =  \frac{dD_{1-}(s)}{ds} \Big| _{s=s_r} = - \omega _r \, ,
\end{eqnarray}
we finally find that
\begin{eqnarray}\label{eq:DA3}
D_A^2 = 
\big| D_{1+}(s_e) D_{1-} (s_e) \big| \, \big( 1 - \frac{\omega _{pr}^2}{\omega _r^2} \big)
\, .
\end{eqnarray}
We have chosen the opening angle in (\ref{eq:vertex}) such that in vacuum we have 
just $D_A^2 = \big| D_{1+}(s_e) D_{1-} (s_e) \big|$.

The \emph{corrected luminosity distance} $D_{L , \mathrm{corr}}$ is defined as
\begin{eqnarray}\label{eq:DLc}
D_{L,\mathrm{corr}} ^2  = \frac{\text{cross-sectional  area  at  the  receiver}}{\text{solid  angle  at  the  emitter}}
\, .
\end{eqnarray}
If a light source isotropically emits photons at a known rate, the number flux 
arriving on the Earth would give us directly its corrected luminosity distance.
We may calculate it in analogy to the area distance, but now with a bundle that has a vertex at
the emitter,
\begin{eqnarray}\label{eq:vertex2}
\boldsymbol{D}{}_2 (s) \big| _{s=s_e} = \boldsymbol{0} \, , \quad
\frac{d}{ds} \boldsymbol{D}{}_2 (s) \big| _{s=s_e}  = \omega _e  \boldsymbol{1} 
\end{eqnarray}
where $\omega _e$ is the frequency of the light ray at the emission event as measured by the observer with 
4-velocity $U_e$. The result is 
\begin{eqnarray}\label{eq:DLc2}
D_{L,\mathrm{corr}} ^2 = 
\big| D_{2+}(s_r) D_{2-} (s_r) \big| \, \big( 1  - \frac{\omega _{pe}^2}{\omega _e ^2 } \big) 
\end{eqnarray}
where $\omega _{pe}$ is the plasma frequency at the emission event.

The (uncorrected) \emph{luminosity distance} $D_L$ differs from $D_{L, \mathrm{corr}}$ by a 
redshift factor, 
\begin{eqnarray}\label{eq:DL}
D_L =\frac{\omega _e}{\omega _r} D_{L,\mathrm{corr}} 
\, ,
\end{eqnarray}
i.e., it is related to the energy flux, rather than to the number flux of photons, that 
arrives at the receiver. The luminosity distance could be directly observed 
if we had \emph{standard candles}, i.e., light sources that isotropically emit photons of 
a certain frequency at a known rate. Then the energy flux from such an object that arrives 
on the Earth would give us directly its corrected luminosity distance. As there are light
sources that come quite close to being standard candles, e.g. supernovae of type Ia, the  
luminosity distance may be viewed as a quantity that can be measured reasonably well.

Upon inserting (\ref{eq:DA3}) and (\ref{eq:DLc2}) into (\ref{eq:DL}), the
reciprocity law (\ref{Rezi mit D+D-}) yields
\begin{eqnarray}\label{eq:DADL}
D_L =\frac{\omega _e}{\omega _r} \, 
\frac{\sqrt{\omega _e^2- \omega _{pe}^2}}{\sqrt{\omega _r^2- \omega _{pr}^2}}
\, D_A 
\, .
\end{eqnarray}
In the well-known vacuum case, $D_L$ and $D_A$ are related by a redshift factor squared.
In a plasma, we see that only one of these two factors is a frequency ratio; the other one
is a wavelength ratio. This reveals a fundamental difference between the 
vacuum and the plasma case. While in the vacuum case we can ignore the difference between 
a thermal and a monochromatic light source, this is no longer the case in the plasma.  
In the vacuum case all frequencies behave in the same way and we can transfer the results 
for monochromatic light sources to those of arbitrary light sources. The vacuum 
version of (\ref{eq:DADL}) is thus usually written for the {\em bolometric}
luminosity distance. In the plasma case, however, we have to specify the frequency, i.e.,
$D_L$ is to be understood as based on a monochromatic luminosity.

We repeat that the given formulas for $D_A$ and $D_L$ in a 
plasma are valid only if both $U_e$ and $U_r$ are orthogonal to the chosen Sachs basis. 
For observers with other 4-velocities, frequencies, angles and cross-sectional areas have
to be transformed with the corresponding formulas from special relativity.

\section{Example: Spatially homogeneous plasma in a Robertson-Walker spacetime}\label{subsec:example}

We specify the spacetime metric to the Robertson-Walker case,
\begin{eqnarray}\label{eq:RW}
g_{\mu \nu} dx^{\mu} dx^{\nu} = - dt^2 + a(t)^2 \bigg( \frac{dr^2}{1-k \, r^2} + r^2
\big( d \vartheta ^2 + \mathrm{sin} ^2 \vartheta \, d \varphi ^2 \big) \bigg)
\, ,
\end{eqnarray}
where $a(t)>0$ is the scale factor and $k$ takes the value 1, 0 or $-1$ depending on whether
the spatial curvature is positive, zero or negative. We assume that the plasma density is spatially
homogeneous, i.e., that $\omega _p$ is a function only of $t$. Then the Hamiltonian
(\ref{eq:Hpl}) for light rays in the plasma reads
\begin{eqnarray}\label{eq:HRW}
H(x,p) = \frac{1}{2} \bigg( - p_t^2 + \frac{1- k \, r^2}{a(t)^2} \,  p_r^2 + \frac{1}{a(t)^2 r^2}
\Big( p_{\vartheta}^2 + \frac{p_{\varphi}^2}{\mathrm{sin} ^2 \vartheta} \Big)  + \omega _p (t) ^2 \bigg) \, .
\end{eqnarray}
When discussing light bundles in this spacetime, because of the symmetry it suffices to 
restrict to the case that the central ray is radial, $d \vartheta / ds = d \varphi /ds =0$.
For such a ray, Hamilton's equations (\ref{H-Gleichungen}) read 
\begin{eqnarray}
\label{eq:HRW1}
\frac{dt}{ds} = - p_t \, ,
\\
\label{eq:HRW2}
\frac{dr}{ds} = \frac{(1-k \, r^2)}{a(t)^2} \, p_r \, ,
\\
\label{eq:HRW3}
\frac{dp_t}{ds} =  \frac{1}{a(t)^3} \frac{da(t)}{dt} \big( 1-k \, r^2 \big) p_r^2  
- \omega _p(t) \frac{d \omega _p(t)}{dt} \, ,
\\
\label{eq:HRW4}
\frac{dp_r}{ds} = \frac{k \, r \, p_r^2}{a(t)^2} \, ,
\\
\label{eq:HRW5}
- p_t^2 + \frac{1-k \, r^2}{a(t)^2} \,  \, p_r^2 + \omega _p (t) ^2 = 0 \, .
\end{eqnarray}
The frequency of the light ray with respect to the observer field $U = \partial _t$ is
\begin{eqnarray}\label{eq:omegaRW}
\omega = - p_{\mu} \delta ^{\mu}_t= - p_t \, .
\end{eqnarray}
Dividing (\ref{eq:HRW3}) by (\ref{eq:HRW1}) and inserting (\ref{eq:HRW5}) yields, after some
rearrangements, the result that 
\begin{eqnarray}\label{eq:C}
C = a(t) \sqrt{\omega(t)^2 - \omega _p (t) ^2}
\end{eqnarray}
is constant along the ray. 

This gives us the redshift law in the Robertson-Walker
universe with a plasma, 
\begin{eqnarray}\label{eq:RWredshift2}
\frac{\sqrt{\omega _e ^2 - \omega _p (t_e) ^2}}{\sqrt{\omega _r ^2 - \omega _p (t_r) ^2}}
= \frac{a(t_r)}{a(t_e)} \, .
\end{eqnarray}
This demonstrates that the wavelength ratio is given in the plasma by the same 
formula as in vacuum,
\begin{eqnarray}\label{eq:RWredshft3}
\frac{\lambda _r}{\lambda _e}
= \frac{a(t_r)}{a(t_e)} = 1 + z\, ,
\end{eqnarray}
but that the frequency ratio is not, recall (\ref{eq:zaZ}). 
If we introduce the \emph{Hubble function} $H(t)$ by the usual equation
\begin{eqnarray}\label{eq:Hubble}
H(t) = \dfrac{1}{a(t)} \dfrac{da(t)}{dt} \, ,
\end{eqnarray}
the wavelength redshift $z$ varies along the ray according to the familiar law
\begin{eqnarray}\label{eq:dzdt}
\dfrac{dz}{dt} =  -(1+z) H(t)
\end{eqnarray}
where $t_r$ is kept fixed and $t=t_e$ parametrises the ray.

\begin{figure}
\includegraphics[width = 0.5 \linewidth]{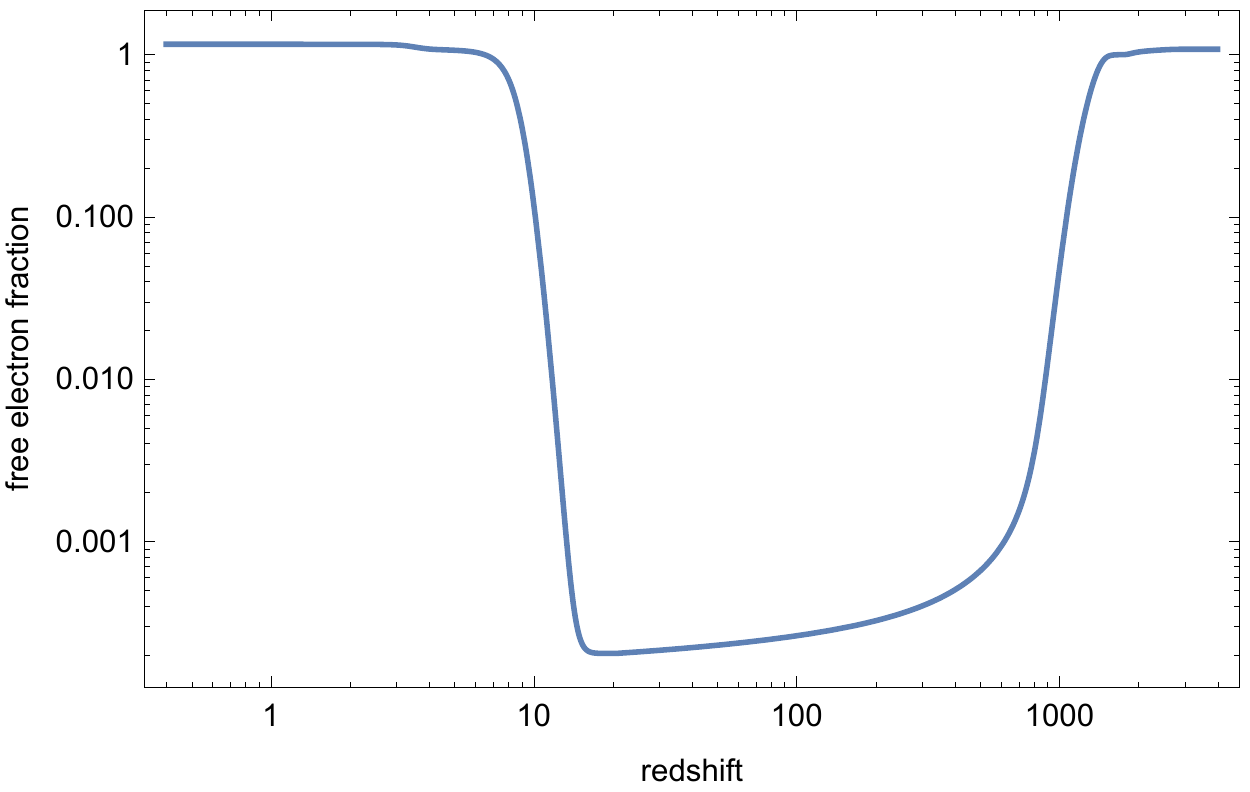}
\includegraphics[width = 0.5 \linewidth]{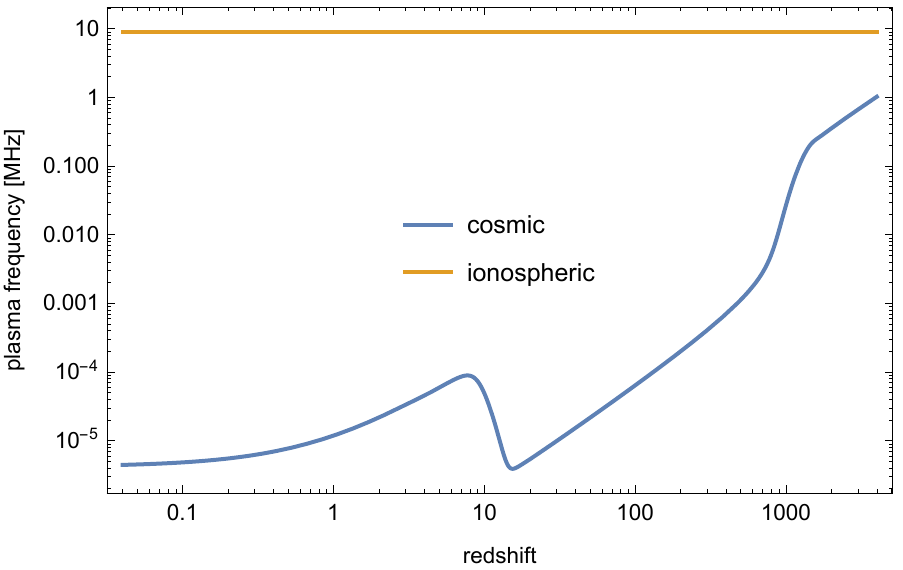}
\caption{{\em Left:} The free electron fraction as a function of cosmological redshift $z$. After the formation of the first atoms 
at $z \sim 1000$ there remains a residual ionisation until the Universe is completely reionised at $z \sim 10$. The effects
of Helium formation and reionisation can be seen as tiny features imprinted on the early and late time plateaus of $x_e$. 
{\em Right:} The cosmic plasma frequency as 
a function of cosmological redshift $z$ and compared to the plasma frequency of Earth's ionosphere. \label{cosmic_history}}
\end{figure}

We can estimate the importance of the difference between the wavelength redshift $z$ and the 
frequency redshift $Z$ by evaluating the plasma frequency as a function of cosmological (wavelength) 
redshift $z$. This is easily done adopting the ratio of free electron number density compared with the 
hydrogen nuclei number density, $x_e$, for the best-fit cosmology
provided by the analysis of the Planck team \cite{Planck2015CosmologicalParameters}. 
Then in (\ref{eq:fp}) we can write $n_e(z) = n_B(0) (1+z)^3 (1 - Y_p) x_e(z)$, where $n_B(0)$ denotes the number 
density of  baryons in the Universe, $Y_p$ is the helium mass fraction and corrects for the effects from helium. The 
factor $(1+z)^3$ accounts for the dilution of the number density due to the expansion. $x_e(z)$ and $f_p(z)$ 
are plotted in Fig.~\ref{cosmic_history}. One can clearly see that the cosmic plasma frequency is well below 
$10$ MHz, the typical plasma frequency of Earth's iononsphere, for the photon transparent Universe. This implies that 
the observable frequency dependent difference (\ref{eq:zZ}),
\begin{eqnarray}\label{eq:Z-z}
Z(z,\omega_r) - z \approx \frac{1}{2\omega_r^2}\left(\frac{\omega^2_p(z)}{(1+z)} 
- (1+z) \omega^2_p(0)\right) \, ,
\end{eqnarray}  
is tiny.

We will now specify the Sachs equations (\ref{S_IPlasma}) and (\ref{S_IIPlasma}) to
the case of a homogeneous plasma on a Robertson-Walker spacetime.
With the help of the constant of motion $C$ we may rewrite the tangent vector to the
light ray as 
\begin{eqnarray}\label{eq:KRW}
\fl \quad
K = K^{\mu} \partial _{\mu} = \frac{dt}{ds} \partial _t + \frac{dr}{ds} \partial _r 
=
\sqrt{\omega _p(t)^2 + \frac{C^2}{a(t)^2}} \, \partial _t \pm \sqrt{1-kr^2} \, \frac{C}{a(t)^2} \, \partial _r \, .
\end{eqnarray}
$C$ is determined if we know the frequency at one particular event along the light ray.
In the vacuum case, $\omega _p =0$, choosing different values for the constant of 
motion has the only effect of choosing different affine parameters for the light ray.

For writing the Sachs equation in the Robertson-Walker universe with a plasma, we
have to choose a Sachs basis along our radial light ray. In the case at hand, a 
natural choice is
\begin{eqnarray}\label{eq:tERW}
\widetilde{E}{}_1 = \frac{1}{\omega _p(t) \, a(t) \, r} \, \partial _{\vartheta} \, , \quad 
\widetilde{E}{}_2 = \frac{1}{\omega _p(t) \, a(t) \, r \, \mathrm{sin} \, \vartheta} \, \partial _{\varphi} 
\, .
\end{eqnarray}
It is easy to see that these two vector fields are, indeed, orthonormal with respect to the
metric $\tilde{g}{}_{\mu \nu} = \omega _p^2 g_{\mu \nu}$. Moreover, with a bit of algebra
one can verify that they satisfy $\widetilde{\nabla} _{\widetilde{K}} \widetilde{E}{}_i = 0$ .
The rescaled Sachs basis (\ref{eq:Epl}) is then obviously given by
\begin{eqnarray}\label{eq:ERW}
E_1 = \frac{1}{a(t) \, r} \, \partial _{\vartheta} \, , \quad 
E_2 = \frac{1}{a(t) \, r \, \mathrm{sin} \, \vartheta} \, \partial _{\varphi} 
\, .
\end{eqnarray}
With $K$, $E_1$ and $E_2$ known we may evaluate the Sachs equations
(\ref{S_IPlasma}) and (\ref{S_IIPlasma}). We just have to calculate the Weyl tensor, the
Ricci tensor and the Ricci scalar of the Robertson-Walker metric (\ref{eq:RW}) which is
an elementary textbook matter. The Weyl tensor vanishes, the non-zero components
of the Ricci tensor are  
\begin{eqnarray}\label{eq:RicciRW}
R_{tt} = \dfrac{3}{a(t)} \, \frac{d^2 a(t)}{dt^2} \, ,
\\
R_{rr} = - \dfrac{1}{(1-k \, r^2)} \,  \left(
2 \, k + 2 \,  \Big( \frac{da(t)}{dt} \Big) ^2
+ a(t) \, \frac{d^2a(t)}{dt^2}  \right) \, ,
\\
R_{\varphi \varphi} = \mathrm{sin}^2 \vartheta \, R_{\vartheta \vartheta}
= \mathrm{sin}^2 \vartheta \, r^2  \big( 1-k \, r ^2 \big) R_{rr}  \, ,
\end{eqnarray}
and the Ricci scalar reads
\begin{eqnarray}\label{eq:RRW}
\mathcal{R} =
- \dfrac{6}{a(t)^2} \, \left( k +  \Big( \frac{da(t)}{dt} \Big) ^2 
+ a(t) \frac{d^2a(t)}{dt^2}  \right) \, .
\end{eqnarray}
The Sachs equations (\ref{S_IPlasma}) and (\ref{S_IIPlasma}) read
\begin{eqnarray}\label{eq:SIRW}
\frac{d \rho}{ds} + \rho ^2 + | \sigma | ^2 = -
\frac{C^2}{a(t)^4} \bigg( k + \Big( \frac{da(t)}{dt} \Big) ^2 - a(t) \frac{d^2a(t)}{dt^2} \bigg)
\\
\nonumber
\qquad \quad 
-
\frac{\omega_p (t)^2}{a(t)^2} \bigg( k + \Big( \frac{da(t)}{dt} \Big) ^2 \bigg) \\
\nonumber 
\qquad \quad
- 
\frac{1}{2} \, \frac{d^2 \omega _p (t)^2}{dt^2}  - \Big( \frac{d \omega_p(t)}{dt}\Big)^2 
- 2 \frac{1}{a(t)} \frac{d a(t)}{dt} \omega_p(t) \frac{d \omega_p(t)}{d t}\, .
\end{eqnarray}
\begin{eqnarray}\label{eq:SIIRW}
\frac{d \sigma}{ds} + 2 \, \theta \,  \sigma = 0 \, .
\end{eqnarray}
From  (\ref{eq:SIIRW}) we read that the plasma does not produce shear, i.e.,
if the shear vanishes at one point, then it vanishes along the entire ray.
For a bundle with $\sigma =0$ and 
$\Omega=0$, the Sachs equations and (\ref{eq:Dpm}) reduce to 
\begin{eqnarray}\label{eq:SachsRW}
\frac{d \theta}{ds} + \theta^2 = 
- \frac{C^2}{a(t)^4} \bigg( k + \Big( \frac{da(t)}{dt} \Big)^2 - a(t) \frac{d^2a(t)}{dt^2} \bigg)
\\
\nonumber
\qquad \qquad
-
\frac{\omega_p (t)^2}{a(t)^2} \bigg( k + \Big( \frac{da(t)}{dt} \Big) ^2 \bigg) \\
\nonumber 
\qquad \qquad
- 
\frac{1}{2} \, \frac{d^2 \omega _p (t)^2}{dt^2}  - \Big( \frac{d \omega_p(t)}{dt}\Big)^2 
- 2 \frac{1}{a(t)} \frac{d a(t)}{dt} \omega_p(t) \frac{d \omega_p(t)}{d t}\, .
\end{eqnarray}
\begin{eqnarray}\label{eq:DpmRW}
\frac{d D_{\pm}}{ds} = D_{\pm} \theta \, .
\end{eqnarray}
Differentiating (\ref{eq:DpmRW}) with respect to $s$ and inserting (\ref{eq:SachsRW})
yields 
\begin{eqnarray}\label{eq:focusRW}
\frac{1}{D_{\pm}} \, \frac{d ^2 D_{\pm}}{ds ^2} = 
- \frac{\big( \omega ^2 - \omega _p (t) ^2 \big)}{a(t)^2} \,
 \bigg( k + \Big( \frac{da(t)}{dt} \Big) ^2 
 - a(t) \frac{d^2a(t)}{dt^2} \bigg)
\\
\nonumber
\qquad \qquad
-
\frac{\omega_p (t)^2}{a(t)^2} \bigg( k + \Big( \frac{da(t)}{dt} \Big) ^2 \bigg) 
\\
\nonumber
\qquad \qquad
- 
\frac{1}{2} \, \frac{d^2 \omega _p (t)^2}{dt^2} - \Big( \frac{d \omega_p(t)}{dt}\Big)^2 
- 2 \frac{1}{a(t)} \frac{d a(t)}{dt} \omega_p(t) \frac{d \omega_p(t)}{d t}\, ,
\end{eqnarray}
where we have inserted the constant of motion $C$ from (\ref{eq:C}). This equation
has to be viewed in conjunction with (\ref{eq:HRW1}), which relates the 
parameter $s$ to the Robertson-Walker time coordinate $t$, and (\ref{eq:HRW3}),
which tells how the frequency changes along the ray. If the time-dependence of the
scale factor $a(t)$ and of the plasma frequency $\omega _p (t)$ has been specified,
integration of this system of equations determines the influence of the plasma on 
the focussing properties and, thus, on the distance measures in the Robertson-Walker 
universe. 

It is convenient to replace the affine parameter $s$ by the cosmological redshift $z$, which is 
done by means  of 
\begin{eqnarray}
\frac{d z}{d s} = \frac{d z}{d t} \frac{d t}{d s} = - (1+z) H(z) \omega(z),
\end{eqnarray}
where we used (\ref{eq:HRW1}), (\ref{eq:omegaRW}) and (\ref{eq:dzdt}) 
in the last step. The Hubble function $H$ and the frequency $\omega$ are now viewed as 
functions of the redshift $z$. In order to get rid of the second 
derivative with respect to $s$ we need to evaluate $d\omega/ds$, which is done by means of 
(\ref{eq:HRW3}),   (\ref{eq:omegaRW})  and (\ref{eq:HRW5}). We also replace all derivatives with respect to 
cosmic time by a derivative with respect to cosmological redshift and obtain after a lengthy calculation
\begin{eqnarray}
\label{eq:evolutionDofz}
\lefteqn{D_\pm'' + \left[\frac{2}{1+z} + \frac{H'}{H} - \frac{1}{\omega^2} \left(\frac{\omega_p^2}{(1+z)^2} - \omega_p 
\omega_p'\right)\right] D_\pm' =} \qquad \nonumber \\
& - & \left[\frac{H'}{(1+z) H} + \frac{k}{a_0^2 H^2} + 
 \frac{\omega_p^2}{\omega^2}\left( \frac{1}{(1+z)^2} - \right. \right. \\ 
&&  \quad \left. \left. \frac{H'}{(1+z) H} + \left(\frac{H'}{H} - \frac{1}{1+z} \right)\frac{\omega_p'}{\omega_p} + 
\frac{2 (\omega_p')^2}{\omega_p^2} + \frac{\omega_p''}{\omega_p}\right) \right] D_\pm, \nonumber
\end{eqnarray}
where a prime denotes a derivative with respect to redshift
and $a_0=a(t_r)$ denotes the value of the scale factor at the reception event
(``today").


For vanishing plasma frequency ($\omega _p =0$) and a spatially flat 
Universe ($k=0$), we recover from this equation the well known result 
$D_\pm(z) = 1/(1+z) \int_0^z d\bar{z}/H(\bar{z})$ if we choose the initial conditions 
appropriately. In this paper we will not attempt to integrate (\ref{eq:evolutionDofz}) 
with a non-vanishing plasma density which, in general, will have to be done numerically.
As an illustration of our general results, we will be satisfied by deriving the influence 
of the plasma on the distance-redshift relation for small redshift, i.e., on the Hubble 
law, in a Robertson-Walker universe. 

Taylor expansion of the function $D_{\pm} (z)$ about the parameter value $z(s_r) = 0$, where 
the light ray meets the receiver, gives
\begin{eqnarray}\label{eq:TaylorD}
D_{\pm} (z) = D_{\pm}(z=0) + D_{\pm}'(z=0) z  + \frac{1}{2} D_\pm''(z=0) z^2 + \dots 
\end{eqnarray}
Imposing the vertex condition (\ref{eq:vertex}) requires 
\begin{eqnarray}\label{eq:vertRW}
& & D_{\pm}(z= 0) = D_{\pm} (s_r) = 0 \, , \\
& & D_{\pm}'(z= 0) = \frac{d s}{dz} (z=0) \frac{dD_{\pm}}{ds} (s_r) = \frac{1}{H_0},
\end{eqnarray}
where $H_0=H(z=0)$ denotes the Hubble constant, 
i.e. the present day value of the Hubble expansion rate.
We further use (\ref{eq:evolutionDofz}) at $z=0$, which gives
\begin{eqnarray}
D_\pm''(z = 0) = - \left[2 + \frac{H'}{H} - \frac{1}{\omega^2} \left(\omega_p^2 - \omega_p 
\omega_p'\right)\right](z=0) \frac{1}{H_0}. 
\end{eqnarray} 
It is convenient to replace the derivatives of the Hubble expansion rate by the deceleration parameter 
$q_0 = (H'/H - 1)(z = 0)$. We find,
\begin{eqnarray}\label{eq:TaylorD2}
D_{\pm} (z) = \frac{z}{H_0} - \frac{z^2}{2 H_0} 
\left[3 + q_0 - \frac{1}{\omega_r^2} \left(\omega_{pr}^2 - \omega_{pr} 
\omega_{pr}' \right)\right] + \dots 
\end{eqnarray}
Here and in the following, we write $\omega _{pr}'$ for 
$\omega _p ' (z=0)$. According to (\ref{eq:DA3}), (\ref{eq:TaylorD2}) gives us the area distance 
\begin{eqnarray}\label{eq:DAsRW}
D_A(z, \omega_r) = 
\sqrt{1 - \frac{\omega _{pr}^2}{\omega_r ^2}}  \frac{z}{H_0}
\left[ 1  + \frac{z}{2} \left(- 3 - q_0 + \frac{\omega_{pr}^2}
{\omega_r^2} - \frac{\omega_{pr} \omega_{pr}'}{\omega_r^2}\right) + \dots \right] .
\end{eqnarray}
The leading term is the well known Hubble law modified by the square root in front of the term linear in redshift.
As the Hubble law is usually formulated for standard candles and not for standard rulers, we can also write it for the 
luminosity distance, by applying the modified reciprocity law
\begin{eqnarray}
D_L(z, \omega_r) = (1+z) (1 + Z(z,\omega_r)) D_A(z, \omega_r).  
\end{eqnarray} 
In order to obtain a consistent Taylor expansion of the luminosity distance up to second order in redshift, 
we Taylor expand
\begin{eqnarray}
(1+z) (1 + Z(z,\omega_r)) = (1+z)^2 \sqrt{1 + \frac{1}{\omega_r^2}\left(\frac{\omega_p^2(z)}{(1+z)^2} - \omega_{pr}^2 \right)} 
\end{eqnarray} 
up to linear order in $z$, which gives
\begin{eqnarray}
(1+z) (1 + Z(z,\omega_r)) = 1 + \left(2 - \frac{\omega_{pr}^2}{\omega_r^2}+ \frac{\omega_{pr} \omega_{pr}'}{\omega_r^2} \right)z + \dots
\end{eqnarray} 
Hence, we find
\begin{eqnarray}\label{eq:DLsRW}
D_L(z, \omega_r) = \sqrt{1 - \frac{\omega _{pr}^2}{\omega_r ^2}}  \frac{z}{H_0}\left[ 1  + \frac{z}{2} \left(1 - q_0 - 
\frac{\omega_{pr}^2}{\omega_r^2} +\frac{\omega_{pr} \omega_{pr}'}{\omega_r^2}\right) + \dots \right] .
\end{eqnarray} 
Comparison of (\ref{eq:DAsRW}) and (\ref{eq:DLsRW}) shows that for the \emph{linear} 
Hubble law it makes no difference if we use $D_A$ or $D_L$. At this order, the 
plasma modifies the vacuum Hubble law by the factor $\sqrt{1 - \omega _{pr}^2 / \omega_r ^2}$.
As this factor is smaller than 1, the effect of the plasma is that light sources at 
a given (area or luminosity) distance appear redder than light sources at the
same (area or luminosity) distance on the same spacetime in vacuum. To estimate the factor 
$\sqrt{1- \omega _{pr}^2/\omega _r^2}$ numerically, we read from Fig. \ref{cosmic_history} that 
at present in our Universe $\omega _{pr} < 10^{-5} \mathrm{MHz}$ while the ionosphere limits us to
$\omega _r > 10 \, \mathrm{MHz}$, hence $\sqrt{1- \omega _{pr}^2/\omega _r^2}
= 1- \varepsilon$ where $\varepsilon < 10^{-12}$. This demonstrates
that the deviation from vacuum is unmeasurably small as long as
we do not have a radio telescope, or better an array of radio
telescopes, in space that could operate at frequencies considerably
below 10 MHz. We have mentioned already in the introduction that
there are some plans for such arrays.
 
Note that we have not used Einstein's field equation. Our result is independent of
whether or not the plasma is self-gravitating and no assumption was made on how the 
scale factor depends on time.   

\section{Conclusions}\label{sec:clonclusions}

The purpose of this work was to discuss the geometry of light bundles
in a non-magnetised cold plasma on an arbitrary spacetime. This is of
relevance for calculating the effects of such a medium on image 
deformation and on distance measures. In fact most of the baryonic matter in the 
Universe is in the aggregate state of a plasma until today. 

In comparison to the vacuum case, the most important difference is in the fact that the shape and the size 
of the cross-section of  a bundle are no longer independent of the observer,
i.e., in a plasma there is no analogue of the Sachs theorem. 
Based on a modified definition of a Sachs basis, we have derived modified Sachs
equations for the shear, the expansion and the twist of a light bundle propagating in a cold plasma. 
These equations allowed us to derive a modified reciprocity
theorem (Etherington law) which relates luminosity distance to area distance.
In vacuum these two distance measures are related by a redshift factor squared.
A similar law holds in a plasma; however, one of the redshift factors is a
frequency ratio whereas the other one is a wavelength ratio. The fact that 
we have to distinguish between these two types of redshift factors is 
another important difference to the vacuum case.

In view of applications to astrophysics or 
cosmology, it seems to us that all plasma effects are negligibly small in the optical frequency 
range but some of them may be observable with radio signals.  

As an illustration of our general results, we have considered a homogeneous
plasma on a Robertson-Walker spacetime. Throughout the paper, and also in
this example, we have not used Einstein's field equation so that the results
apply to a self-gravitating plasma and equally well to a ``test plasma'' on 
a given spacetime background. In the Robertson-Walker case, we have 
investigated the influence of the plasma on distance measures and on
the distance-redshift relation. We believe that these results are of 
some interest from a conceptual point of view, although the difference to
the vacuum case is too small for being observed with existing instruments. 
However, the real Universe does contain structure and  the 
magnitude of the expected plasma effect does not only depend on the plasma frequency and 
its time derivatives, but also its spatial variations.

Our general results may be applied to other examples, e.g. to a 
spherically sym\-met\-ric plasma around a Schwarzschild black hole. In
this case the plasma would have an influence on image deformation, in
particular on the size of Einstein rings, and on the magnification of images.
We are planning to discuss this example in another paper. Other interesting applications 
might be weak and strong lensing of radio galaxies by galaxy clusters, whose baryonic matter content 
is known to be dominated by the cluster plasma. We would expect the achromatic gravitational lensing at 
radio frequencies to be modified by chromatic corrections from plasma effects.

\section*{Acknowledgements}

We would like to thank Isabel Oldengott for providing the free electron fraction $x_e(z)$ and we thank 
Marcus Br\"uggen, Walter Pfeiffer and Joris Verbiest for valuable discussions. We gratefully acknowledge 
support from the DFG within the Research Training Group 1620 ``Models of Gravity''. 


\section*{References}

\bibliographystyle{iopart-num}

\end{document}